\documentclass[twocolumn, twocolappendix]{aastex631}

\usepackage{amsmath}
\usepackage{amssymb}
\usepackage{graphicx}
\usepackage[normalem]{ulem}

\newcommand{\fNL}{f_\mathrm{NL}}
\newcommand{\fNLloc}{f_\mathrm{NL}^\mathrm{local}}
\newcommand{\fNLeq}{f_\mathrm{NL}^\mathrm{equil}}
\newcommand{\fNLort}{f_\mathrm{NL}^\mathrm{ortho}}

\newcommand{\Om}{\Omega_m}
\newcommand{\Mmin}{M_\mathrm{min}}
\newcommand{\Mmax}{M_\mathrm{max}}

\newcommand{\hMpc}{h\,\mathrm{Mpc}^{-1}}

\newcommand{\kmax}{k_\mathrm{max}}

\def\bs{\mathbf{s}}
\def\cov{\mathbf{C}}

\newcommand{\Planck}{\textit{Planck}}
\newcommand{\Quijote}{\textsc{Quijote}}
\newcommand{\QuijotePNG}{\textsc{Quijote-png}}

\newcommand{\ias}{Universit\'{e} Paris-Saclay, CNRS, Institut d’Astrophysique Spatiale, 91405, Orsay, France}
\newcommand{\cnrs}{Sorbonne Universit\'{e}, CNRS, UMR 7095, Institut d'Astrophysique de Paris, 98 bis bd Arago, 75014 Paris, France}
\newcommand{\cca}{Center for Computational Astrophysics, Flatiron Institute, 162 5th Avenue, New York, NY 10010, USA}
\newcommand{\bologna}{Dipartimento di Fisica e Astronomia, Alma Mater Studiorum - University of Bologna, Via Piero Gobetti 93/2, 40129 Bologna BO, Italy}
\newcommand{\inaf}{INAF - Osservatorio Astronomico di Bologna, Via Piero Gobetti 93/3, 40129 Bologna BO, Italy}
\newcommand{\infn}{INFN - Istituto Nazionale di Fisica Nucleare, Sezione di Bologna, Viale Berti Pichat 6/2, 40127 Bologna BO, Italy
}
\newcommand{\infnPad}{INFN, Sezione di Padova, via Marzolo 8, I-35131, Padova, Italy}
\newcommand{\Trento}{Dipartimento di Fisica, Universit\`a degli Studi di Trento, Via Sommarive 14, I-38123 Povo (TN), Italy}
\newcommand{\mpa}{Max-Planck-Institut f\"ur Astrophysik, Karl-Schwarzschild-Straße 1, 85748 Garching, Germany}
\newcommand{\ICC}{ICC, University of Barcelona, IEEC-UB, Martí i Franquès, 1, E-08028 Barcelona, Spain}
\newcommand{\ICREA}{ICREA, Pg. Lluís Companys 23, Barcelona, E-08010, Spain}
\newcommand{\Galilei}{Dipartimento di Fisica e Astronomia “G. Galilei”,Università degli Studi di Padova, via Marzolo 8, I-35131, Padova, Italy}
\newcommand{\uwc}{Department of Physics and Astronomy, University of the Western Cape, Cape Town 7535, South Africa}
\newcommand{\princeton}{Department of Astrophysical Sciences, Princeton University, 4 Ivy Lane, Princeton, NJ 08544 USA}

\begin{document}
\title{Quijote-PNG: The Information Content of the Halo Mass Function}

\author[0000-0002-2133-6254]{Gabriel Jung}
\affiliation{\ias}

\author[0000-0002-1370-0546]{Andrea Ravenni}
\affiliation{\Galilei}
\affiliation{\infnPad}

\author{Marco Baldi}
\affiliation{\bologna}
\affiliation{\inaf}
\affiliation{\infn}

\author{William R Coulton}
\affiliation{\cca}

\author{Drew Jamieson}
\affiliation{\mpa}

\author{Dionysios Karagiannis}
\affiliation{\uwc}

\author{Michele Liguori}
\affiliation{\Galilei}
\affiliation{\infnPad}
\affiliation{\Trento}

\author{Helen Shao}
\affiliation{\princeton}

\author{Licia Verde}
\affiliation{\ICREA} 
\affiliation{\ICC}

\author{Francisco Villaescusa-Navarro}
\affiliation{\cca}
\affiliation{\princeton}

\author{Benjamin D. Wandelt}
\affiliation{\cnrs}
\affiliation{\cca}

\begin{abstract}

We study signatures of primordial non-Gaussianity (PNG) in the redshift-space halo field on non-linear scales using a combination of three summary statistics, namely, the halo mass function (HMF), power spectrum, and bispectrum. The choice of adding the HMF to our previous joint analysis of the power spectrum and bispectrum is driven by a preliminary field-level analysis, in which we train graph neural networks on halo catalogues to infer the PNG $\fNL$ parameter. The covariance matrix and the responses of our summaries to changes in model parameters are extracted from a suite of halo catalogues constructed from the \QuijotePNG\ N-body simulations. We consider the three main types of PNG: local, equilateral and orthogonal. Adding the HMF to our previous joint analysis of power spectrum and bispectrum produces two main effects. First, it reduces the equilateral $\fNL$ predicted errors by roughly a factor $2$, while also producing notable, although smaller, improvements for orthogonal PNG. Second, it helps break the degeneracy between the local PNG amplitude, $\fNLloc$, and assembly bias, $b_{\phi}$, without relying on any external prior assumption. Our final forecasts for PNG parameters are $\Delta \fNLloc = 40$, $\Delta \fNLeq = 200$, $\Delta \fNLort = 85$, on a cubic volume of $1 \left( h^{-1}{\rm Gpc} \right)^3$, with a halo number density of $\bar{n}\sim 5.1 \times 10^{-5} ~h^3\mathrm{Mpc}^{-3}$, at $z = 1$, and considering scales up to $\kmax = 0.5~\hMpc$.

\end{abstract}

\section{Introduction}
\label{sec:introduction}

The presence of a certain degree of non-Gaussianity (NG) in the primordial cosmological perturbation field is a general prediction of both inflationary and other early Universe scenarios. In addition, both the level of the predicted NG signal and the shape of the expected NG signatures are significantly model dependent. This makes primordial non-Gaussianity (PNG) a powerful tool to constrain inflation, or alternative primordial models, and to provide clues about physics at very high energy scales. 

From an observational point of view, the challenging aspect of any PNG analysis is that the expected NG signatures are very small and the optimal statistic that maximizes their signal-to-noise ratio is unknown from low-redshift observables. Indeed, to date there has been no experimental detection of a PNG signal, although significant constraints have been placed using Cosmic Microwave Background (CMB) data; the CMB is an ideal observable for PNG studies, since it formed at early times, when cosmological perturbations were still in the linear regime, hence preserving the statistical features of the primordial fluctuation field. The most precise results currently come from the analysis of \Planck\ CMB data, which produced an upper bound on the level of PNG at roughly less than $0.1\%$ of the amplitude of the Gaussian component of the field \citep{Planck:2019kim}. 

The open question is whether and how we can obtain more stringent PNG constraints---or achieve a detection---with future cosmological observations.
In this respect, it is known that, after \Planck, CMB data have nearly saturated their PNG constraining power, with possible improvements of, at most, a factor of $\sim 2$ for relevant parameters in a majority of scenarios \citep{CORE:2016ymi, Abazajian:2019eic}. It is therefore necessary to explore different observables. Galaxy clustering is a natural candidate for two main reasons. First of all, in the limit of weak PNG, the bispectrum (i.e., the $3$-point function of the Fourier/harmonic modes) of primordial cosmological perturbations contains most of the non-Gaussian information and the three-dimensional galaxy density field contains more bispectrum modes for NG analysis than the two-dimensional CMB map. 
Furthermore, some models---notably, 
 those producing a ``local type" bispectrum, where the signal peaks on squeezed Fourier mode triangles---generate a characteristic scale dependent signature in the galaxy power spectrum on very large scales \citep{Dalal:2007cu, Matarrese:2008nc, Slosar:2008hx, McDonald:2008sc, Giannantonio:2009ak, Desjacques:2010jw}, which can be used to constrain NG.

In both cases, however, there are some important complications to consider. 
As far as bispectrum analysis is concerned, the big caveat is that the additional modes in the Large Scale Structure (LSS) bispectrum are in the non-linear regime. Hence, they present a ``late-time'' component generated by the non-linear gravitational evolution of structures, which is hard to disentangle and much larger than the primordial one. Of course, this late-time 3-point signal is interesting in itself since it carries a lot of information about cosmological parameters and structure evolution \citep{Chang_I, Chang_II}; however, as long as we are focused on PNG, it is  a massive source of contamination, with an amplitude $\sim 1000$ times larger than the primordial signal of interest. 
The scale-dependent power spectrum signature on large scales clearly does not present this problem and was considered for a long time to be a cleaner LSS probe of PNG, although limited to a subset of all possible PNG scenarios. However, a significant issue has recently been pointed out again in this area \citep{Reid:2010vc, Barreira:2020ekm, Barreira:2022sey}, namely, the degeneracy produced by the breaking of the universality relation that was generally used to link the NG galaxy bias parameter $b_{\phi}$ to the linear bias parameter $b_1$. This is due to halo/galaxy assembly bias effects, and, if not addressed in any way, it allows us only to constrain the $b_{\phi} f_{\rm NL}$ combination.

A key objective in cosmological PNG studies is thus developing optimal data analysis strategies to overcome, at least partially, the aforementioned issues. As far as the $b_{\phi}(b_1)$ relation is concerned, an active effort is being put into characterizing it as well as possible via numerical studies of N-body simulations \citep{Barreira:2020ekm, Barreira:2022sey,Lazeyras:2022koc, Sullivan:2023qjr}, in order to produce accurate priors. Another logical line of attack, which we start exploring in this work, is that of going beyond a power spectrum $+$ bispectrum analysis and including extra summary statistics, which could help disentangle the PNG signal from late-time evolution effects. The open question, with no straightforward answer, is of course, which summary statistics are best suited to this purpose? In this paper, we explore the halo mass function (HMF) as an interesting candidate. This choice was not casual but was driven by training graph neural networks to perform field-level likelihood-free inference on halo catalogues from \QuijotePNG\ simulations. The analysis of the outcome of those calculations led us to the conclusion that the model was extracting information from the abundance of halos, as we explain in section\ \ref{sec:GNN}. Therefore, the HMF can be seen as a machine learning-driven statistic that stands ahead of others.

Furthermore, our choice is also justified at a theoretical level, since the HMF has been known for a long time to be sensitive to non-Gaussian initial conditions (ICs), which are able to skew its distribution by changing the abundance of massive halos, and it was proposed as an interesting complementary PNG probe to the bispectrum in a number of papers \citep{Matarrese:2000iz, Sefusatti:2006eu, Grossi:2007ry, Pillepich:2008ka, Desjacques:2008vf, Grossi:2009an, Desjacques:2010nn, LoVerde:2011iz, Palma:2019lpt}. On top of this, a major advantage of the HMF is that it directly depends on the PNG amplitude parameter $f_{\rm NL}$. Therefore, it does not exhibit the $b_{\phi}$--$f_{\rm NL}$ degeneracy that affects the scale-dependent power spectrum signature. 

This work belongs to the \QuijotePNG\ series \citep{Coulton:2022qbc,Jung:2022rtn,Coulton:2022rir,Jung:2022gfa}, where we aim to build a simulation-based pipeline to optimally extract NG information, pushing our analysis to smaller, non-linear scales. This kind of approach is complementary to a perturbation theory-based, likelihood analysis of power spectrum and bispectrum \citep{2021JCAP...05..015M,Cabass:2022ymb,Cabass:2022wjy,DAmico:2022gki}. See also \citet{Giri:2023mpg}, for an alternative simulation-based approach that uses large scale modulation of small scale power.

The paper is structured as follows. In section \ref{sec:simulations} we briefly describe the simulation dataset used in our analysis; in section \ref{sec:GNN} we describe our preliminary field-level analysis; in section \ref{sec:fisher} we recall and summarize the main methodological aspects of our data analysis pipeline to extract relevant summary statistics and compute the corresponding Fisher matrix; section \ref{sec:hmf} is devoted to a specific discussion of the HMF, the main new ingredient with respect to our previous analyses, and of how we extract it from simulations; our numerical Fisher forecasts are described in section \ref{sec:analysis}, where we also discuss the improvements coming from complementing the initial power spectrum $+$ bispectrum analysis with HMF estimates; finally, we draw our conclusions in section \ref{sec:conclusion}.

\section{Simulations}
\label{sec:simulations}

In this work, we use the publicly available halo catalogues derived from the \Quijote\ suite of N-body simulations \citep{Villaescusa-Navarro:2019bje}.\footnote{\url{https://quijote-simulations.readthedocs.io}} These simulations have been produced using the codes \textsc{2LPTIC} \citep{Crocce:2006ve} and \textsc{2LPTPNG} \citep{Scoccimarro:2011pz, Coulton:2022qbc}\footnote{\url{https://github.com/dsjamieson/2LPTPNG}} to generate ICs at $z=127$, \textsc{Gadget-III} \citep{Springel:2005mi} to follow their evolution up to $z=0$ and the friends-of-friends (FOF) algorithm to identify the halos in each simulation \citep{1985ApJ...292..371D}. 

We report the cosmological parameters of these simulations in table\ \ref{tab:quijote}. As described in section\ \ref{sec:fisher}, we use 15,000 simulations of the fiducial cosmology to evaluate covariance matrices, and paired sets of $500$ catalogues where one parameter is displaced by a small step from its fiducial value to compute derivatives with respect to all parameters considered in the analyses. As in \citet{Coulton:2022rir, Jung:2022gfa}, we focus on the cosmological parameters $\left\{\sigma_8, \Om, n_s, h \right\}$\footnote{We do not include $\Omega_{\rm b}$ in the analyses presented here, as it is the parameter that is most affected by the numerical convergence issue mentioned in section~\ref{sec:fisher} and it does not significantly impact the results. Moreover, $\Omega_{\rm b}$ is better constrained by CMB observations.} and PNG amplitudes $\{\fNLloc, \fNLeq, \fNLort \}$, including a simplified bias parameter $\Mmin$ (the minimum mass of halos included in the analysis). To ensure that the IC generation method has not generated unphysical higher-order N-point functions, which could impact the results presented here, we performed further validation of the initial conditions by examining the primordial trispectrum. As is discussed in appendix~\ref{app:IC-tests}, we find no evidence of large, unphysical trispectra in the ICs.

We focus our analyses on redshift $z=1$, for which all power spectra and (modal) bispectra have been computed in \citet{Jung:2022gfa}. Results at lower redshifts, $z=0.5$ and $z=0$, are also shown in appendix~\ref{app:other-redshifts}.

\begin{deluxetable*}{c|c|ccccccccc}[]
\tablecaption{The parameters of the \Quijote~and \QuijotePNG~halo catalogues used in this work.  \label{tab:quijote}}
\tablehead{& $N_\mathrm{sims}$ & $\sigma_8$ & $\Omega_m$ & $\Omega_{\rm b}$ & $n_s$ & $h$ & $\fNLloc$ & $\fNLeq$ & $\fNLort$ & $M_\mathrm{min} (M_\odot/h)$}
\startdata
Fiducial & $15000$ & $0.834$ & $0.3175$ & $0.049$ & $0.9624$ & $0.6711$ & $0$ & $0$ & $0$ & $3.2\times10^{13}$\\
 \hline
$\sigma_8^{+}$ & $500$ & $0.849$ & $0.3175$ & $0.049$ & $0.9624$ & $0.6711$ & $0$ & $0$ & $0$ & $3.2\times10^{13}$\\
$\sigma_8^{-}$ & $500$ & $0.819$ & $0.3175$ & $0.049$ & $0.9624$ & $0.6711$ & $0$ & $0$ & $0$ & $3.2\times10^{13}$\\
$\Omega_m^{+}$ & $500$ & $0.834$ & $0.3275$ & $0.049$ & $0.9624$ & $0.6711$ & $0$ & $0$ & $0$ & $3.2\times10^{13}$\\
$\Omega_m^{-}$ & $500$ & $0.834$ & $0.3075$ & $0.049$ & $0.9624$ & $0.6711$ & $0$ & $0$ & $0$ & $3.2\times10^{13}$\\
$n_s^{+}$ & $500$ & $0.834$ & $0.3175$ & $0.049$ & $0.9824$ & $0.6711$ & $0$ & $0$ & $0$ & $3.2\times10^{13}$\\
$n_s^{-}$ & $500$ & $0.834$ & $0.3175$ & $0.049$ & $0.9424$ & $0.6711$ & $0$ & $0$ & $0$ & $3.2\times10^{13}$\\
$h^{+}$ & $500$ & $0.834$ & $0.3175$ & $0.049$ & $0.9624$ & $0.6911$ & $0$ & $0$ & $0$ & $3.2\times10^{13}$\\
$h^{-}$ & $500$ & $0.834$ & $0.3175$ & $0.049$ & $0.9624$ & $0.6511$ & $0$ & $0$ & $0$ & $3.2\times10^{13}$\\
 \hline
$f_\mathrm{NL}^{\mathrm{local},+}$ & $500$ & $0.834$ & $0.3175$ & $0.049$ & $0.9624$ & $0.6711$ & $+100$ & $0$ & $0$  & $3.2\times10^{13}$\\
$f_\mathrm{NL}^{\mathrm{local},-}$ & $500$ & $0.834$ & $0.3175$ & $0.049$ & $0.9624$ & $0.6711$ & $-100$ & $ 0$ & $0$ & $3.2\times10^{13}$ \\
$f_\mathrm{NL}^{\mathrm{equil},+}$ & $500$ & $0.834$ & $0.3175$ & $0.049$ & $0.9624$ & $0.6711$ & $0$ & $+100$ & $0$ & $3.2\times10^{13}$\\
$f_\mathrm{NL}^{\mathrm{equil},-}$ & $500$ & $0.834$ & $0.3175$ & $0.049$ & $0.9624$ & $0.6711$ & $ 0$ & $-100$ & $0$ & $3.2\times10^{13}$ \\
$f_\mathrm{NL}^{\mathrm{ortho},+}$ & $500$ & $0.834$ & $0.3175$ & $0.049$ & $0.9624$ & $0.6711$ & $0$ & $0$ & $+100$ & $3.2\times10^{13}$\\
$f_\mathrm{NL}^{\mathrm{ortho},-}$ & $500$ & $0.834$ & $0.3175$ & $0.049$ & $0.9624$ & $0.6711$ & $ 0$ & $0$ & $-100$ & $3.2\times10^{13}$ \\
 \hline
$M_\mathrm{min}^{+}$ & $500$ & $0.834$ & $0.3175$ & $0.049$ & $0.9624$ & $0.6711$ & $0$ & $0$ & $0$ & $3.3\times10^{13}$\\
$M_\mathrm{min}^{-}$ & $500$ & $0.834$ & $0.3175$ & $0.049$ & $0.9624$ & $0.6711$ & $0$ & $0$ & $0$ & $3.1\times10^{13}$
\enddata
\end{deluxetable*}

\section{Methods}
\label{sec:method}

\subsection{Field-level analysis}
\label{sec:GNN}

As we discussed in the introduction, the problem of finding an optimal summary statistic that minimizes the error bars on a given cosmological or PNG parameter is unsolved. An alternative to using summary statistics is to perform field-level analysis. The goal with this kind of analysis is to maximize the amount of information that can be extracted without relying on summary statistics. While there are many types of methods to perform such analysis, in our case we made use of graph neural networks (GNNs) \citep{Battaglia_2018}. The advantages of GNNs over other methods are that they 1) do not impose a cut on scales; 2) symmetries (e.g. rotational and translational invariance) can be easily implemented; and 3) can be more interpretable than other methods. Because of this, we decided to train GNNs to perform field-level likelihood-free inference. 

As a starting point, we run 1,000 simulations; each containing $512^3$ particles in a periodic box of size $1~h^{-1}{\rm Gpc}$. Each of those simulations has a different initial random seed but also a different value of $\fNLloc$ in the range $-300$, $+300$. The value of the cosmological parameters was the same in all simulations. We then trained a GNN to perform field-level likelihood-free inference on the value of $\fNLloc$. The architecture and training procedure are the same as those outlined in \citet{Natali_2023, Helen_2022, Pablo_2022}. 

From this exercise, we found that our model was able to infer the value of $\fNLloc$ with an error of $\sigma(\fNLloc) \sim 35$, at $z=0$. In an attempt to understand the behavior of the network, we trained a deep set model \citep{deep_sets} where the only information we made use of the halos was their masses, not their spatial positions. By training such a model, we found that the performance of this model was almost identical to the one of the GNNs. We thus concluded that the network was likely not using the clustering of the halos to perform the inference. Therefore, the network should be using the abundance of halos to infer $\fNLloc$. 

To verify this, we trained a simple model consisting of fully connected layers on the HMF of the halo catalogues from the simulations. We found that this model performed almost as well as the GNN. From this exercise, we reached the conclusion that the HMF is a summary statistic that contains lots of information, likely more than clustering-based statistics, as the GNN did not use those to perform the inference. We emphasize that we trained the GNN using halo catalogues from simulations that only vary $\fNLloc$. Therefore, our results did not account for degeneracies with cosmological parameters that could degrade the constraints, as we shall see below.

This motivated a further analysis, illustrated in the following sections, in which we explicitly extract the power spectrum, bispectrum and HMF from the \Quijote\ dataset, as well as their covariance and response to variations in both cosmological and PNG parameters, in order to perform a full Fisher matrix forecast on non-linear scales. 

\subsection{Fisher information}
\label{sec:fisher}

In this section, we recall the main ingredients of our Fisher analysis pipeline, which was previously used in \citet{Jung:2022gfa}.

The Fisher information matrix, defined as
\begin{equation}
    \label{eq:fisher}
    F_{ij} = \left(\frac{\partial \bar{\bs}}{\partial \theta_i}\right)^\mathrm{T} \cov^{-1}\left(\frac{\partial \bar{\bs}}{\partial \theta_j}\right),
\end{equation}
allows us to estimate the variance, $\sigma^2(\theta_i)=\sqrt{(F^{-1})_{ii}}$, of the optimal unbiased estimator of a given summary statistic $\bs$ with covariance $\cov$ assuming the statistic is Gaussian distributed,\footnote{As verified in \citet{Jung:2022rtn, Jung:2022gfa}, this is a good approximation for the power spectrum and bispectrum} and neglecting the dependence of $\cov$ itself on parameters \citep[][]{Carron:2012pw}.

In this work, both the covariance and derivatives are computed from the simulations described in section~\ref{sec:simulations}. The covariance matrix is evaluated using
\begin{equation}
    \label{eq:covariance}
    \hat{\cov} = \frac{1}{n_r-1} (\bs - \bar{\bs}) (\bs - \bar{\bs})^\mathrm{T},
\end{equation}
where $n_r$ is the number of realizations at fiducial cosmology (15,000 here). Then, to obtain an unbiased estimate of the precision matrix, we apply the Hartlap correction factor \citep{Hartlap:2006kj}
\begin{equation}
    \label{eq:hartlap}
    \cov^{-1} = \frac{n_r-n_s-2}{n_r-1}\hat{\cov}^{-1},
\end{equation}
where $n_s$ is the length of the summary statistic vector $\bs$ (note, however, that this correction is very small here as $n_s \sim 10^{2}$ while $n_r=15000$).

The derivatives are calculated using finite difference,
\begin{equation}
    \label{eq:derivative}
    \frac{\partial \bar{\bs}}{\partial \theta_i} = \frac{\bar{\bs}(\theta_i^\mathrm{fid} + \delta\theta_i) - \bar{\bs}(\theta_i^\mathrm{fid} - \delta\theta_i)}{2\delta\theta_i},
\end{equation}
where we use the sets of 500 simulations where one parameter $\theta_i$ is displaced by $\pm\delta\theta_i$ with respect to its fiducial value. However, it was noticed in \citet{Coulton:2022rir, Jung:2022gfa} that this number of realizations was not sufficient to obtain fully converged derivatives of the halo power spectrum and bispectrum, leading to spuriously low predictions when analyzing jointly cosmological parameters and PNG amplitudes. To overcome this issue, a conservative approach to Fisher matrix computations was developed in \citet{Coulton:2023sfu, Coulton:2022rir}, that is based on computing the Fisher matrix from maximally compressed statistics instead of working with the summary statistics directly.

As shown in \citet{Heavens:1999am, Alsing:2017var}, the compressed quantity defined by
\begin{equation}
    \label{eq:compression}
    \tilde{s}_i = \frac{\partial \bar{\bs}}{\partial \theta_i} \cov^{-1} (\bs - \bar{\bs}),
\end{equation}
conserves all of the statistical information about the parameter $\theta_i$ contained in the data vector $\bs$, if $\bs$ follows a Gaussian likelihood (hence, the same assumption as for the Fisher matrix in eq.~\ref{eq:fisher}). This compression uses the same ingredients as for the Fisher matrix computation (covariance and derivatives of $\bs$), with the addition of the mean $\bar{\bs}$ that is trivial to evaluate from the simulations at fiducial cosmology. Repeating the process for all parameters of interest in $\theta$, one can then compute the Fisher matrix of the compressed statistics $\tilde{\bs}$ by substituting it for $\bs$ in eq.\ \eqref{eq:fisher}. In practice, one has to separate the initial dataset into two subsets. The first is used to perform the compression (i.e.\ compute the derivatives in eq.\ \ref{eq:compression}) and the second is compressed (i.e.\ $\bs$ in eq.\ \ref{eq:compression}) and is then used to calculate derivatives $\partial \tilde{\bs}/\partial \theta_i$ and covariance $\hat{\cov}$ of the compressed statistics, to obtain a conservative estimation of the Fisher matrix. In this work, we use $80\%$ and $20\%$ of the simulations for the two steps respectively, which have been verified to give optimal and numerically stable results. We repeat the procedure for many random splits of the data (between the two steps) and average the results to minimize the intrinsic variance of the method. 

Finally, as shown in \citet{Coulton:2023sfu}, computing the following combination of the standard (overoptimistic) and compressed (conservative) Fisher matrices
\begin{equation}
    \label{eq:fisher-combined}
    F^\mathrm{combined} = G(F^\mathrm{standard}, F^\mathrm{compressed}),
\end{equation}
where $G$ corresponds to the geometric mean defined by 
\begin{equation}
    \label{eq:geometric-mean}
    G(A,B)=A^\frac{1}{2}(A^{-\frac{1}{2}} B  A^{-\frac{1}{2}} )^\frac{1}{2}A^\frac{1}{2},
\end{equation}
gives unbiased estimates of the Fisher error bars with a much smaller number of simulations. An illustration of the different convergences for the three methods is provided in appendix~\ref{app:convergence}.

\subsection{Halo mass function}
\label{sec:hmf}

In addition to the halo power spectrum and bispectrum, we consider the HMF defined as the number of dark matter halos per unit of comoving volume per unit of logarithmic mass bins.

 We measure it in the \Quijote\ simulations using $15$ logarithmic bins corresponding to halo masses $M$ between approximately $2.0 \times 10^{13}$ and $4.6 \times 10^{15}~M_\odot/h$ (note, however, that we do not use the first two bins in the analyses presented in section~\ref{sec:analysis}). To be exact, we use the same binning as in \citet{Bayer:2021iyb}, where the counted halos each contain between $30$ and $7000$ dark matter particles.\footnote{The mass of a halo is given by $M=N m_p$ where $N$ is the number of dark matter particles it contains, and $m_p$ is the mass of a dark matter particle. However, $m_p$ depends on the cosmological parameter $\Om$, which requires the inclusion of the correction term $-\frac{1}{\Om} \frac{\partial\,\mathrm{HMF}}{\partial\,\mathrm{ln}N}$ when computing the derivative $\frac{\partial\,\mathrm{HMF}}{\partial\,\Om}$ \citep[see][for details]{Bayer:2021iyb}. This derivative can also be evaluated by finite difference between bins of $N$.}

\begin{figure}
    \includegraphics[width=0.99\linewidth]{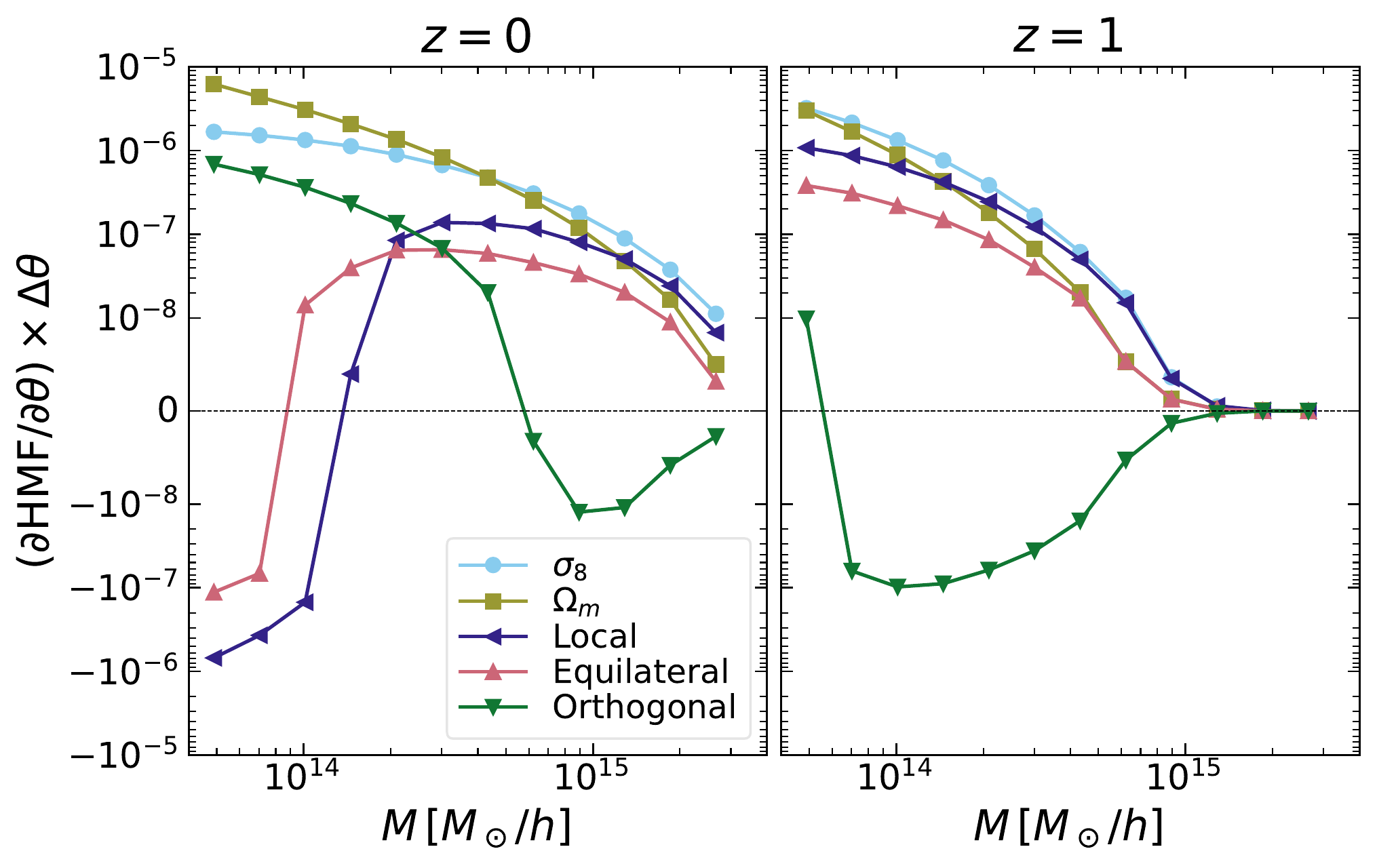}
    \caption{The HMF derivatives with respect to the parameters $\left\{ \sigma_8, \Omega_m,\fNLloc, \fNLeq, \fNLort\right\}$ at $z=0$ and $1$. For internal comparison, the derivative with respect a given parameter $\theta$ is multiplied by the finite difference $\Delta \theta$, used for its numerical estimation (see table \ref{tab:quijote} for details). The vertical scale is logarithmic, except in the range $[-10^{-8},10^{-8}]$, where it is linear. Note that, in some cases, we have a change of sign in the $f_{\rm NL}$ derivatives, implying an opposite effect of PNG on the abundance of high- and low-mass halos, respectively. This is consistent with previous findings in the literature, as pointed out in the main text. The decreasing behaviour of all derivatives at high $M$ is related to the exponential decay of the HMF in this mass range; note that a plot of the logarithmic derivatives would display clear differences between them, also at high $M$. The numerical results displayed here have all been cross-validated in the simulation-independent, halo-model based analysis that we describe in section \ref{sec:bias}.}
    \label{fig:deriv-hmf}
\end{figure}

In figure~\ref{fig:deriv-hmf}, we show the impact of the three shapes of PNG on the HMF. Both the local and equilateral shapes increase the number of massive halos for a positive $\fNL$ value (and decrease it for a negative $\fNL$) and have very degenerate signatures, while for orthogonal PNG it is the opposite. For less massive halos, the effect of PNG changes sign (with the switch occurring for higher masses for orthogonal PNG, which is the only one that appears in the mass range of the plot at $z=1$). This effect was already present on early works on the HMF with PNG simulations \citep[see e.g.][]{LoVerde:2007ri} and is due to the fact that, at fixed $\Omega_m$, more massive halos can only appear at the expense of less massive halos and matter in smaller structures. 

\section{Results}
\label{sec:analysis}

\subsection{Constraints from the HMF}
\label{sec:hmf:constraints}

\begin{figure}
    \includegraphics[width=0.99\linewidth]{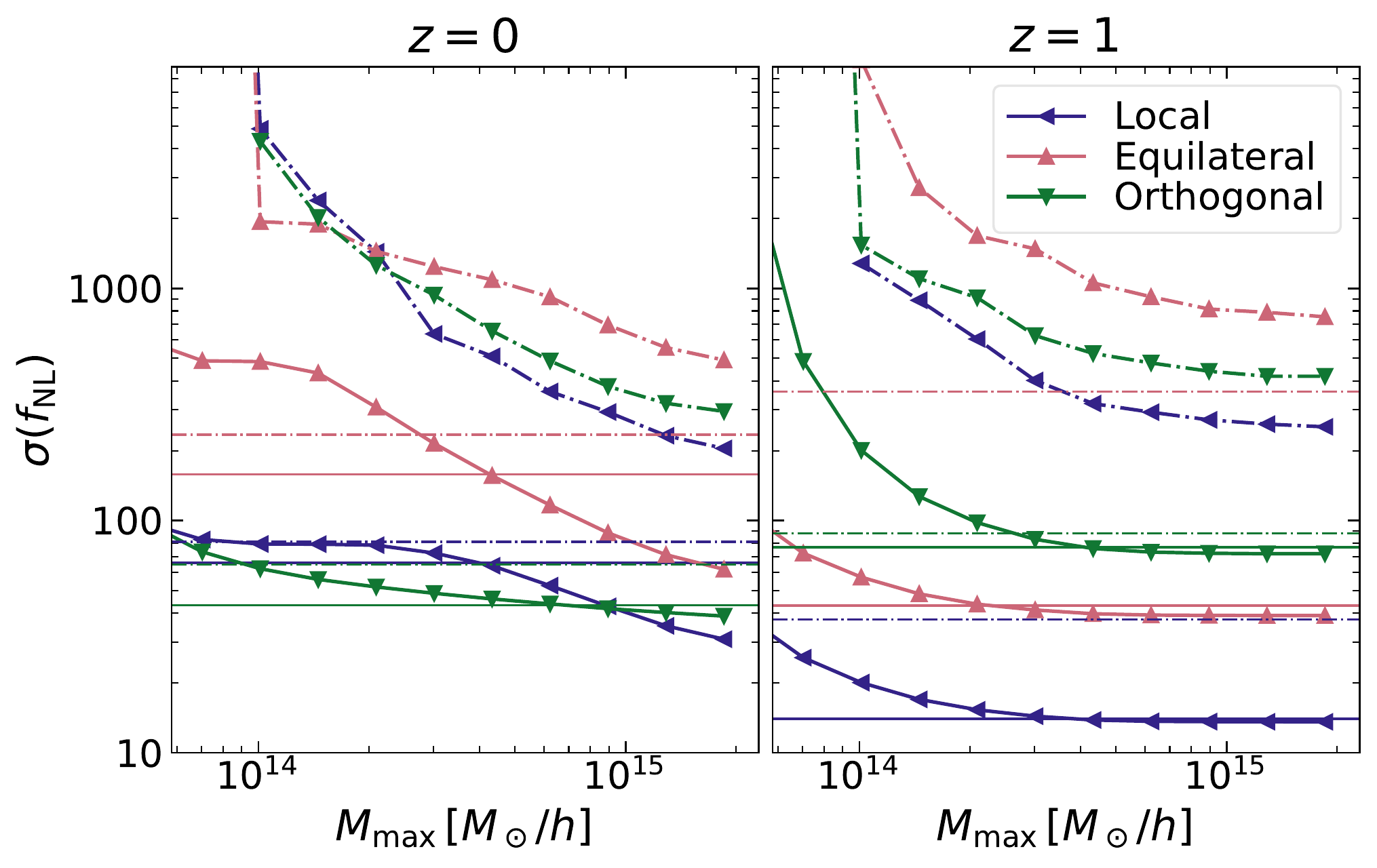}
    \caption{The $1$-$\sigma$ Fisher error bars on $\fNL$ (local, equilateral and orthogonal) from the HMF, as a function of the maximum mass $\Mmax$ of halos considered ($\Mmin \sim 4.1\times 10^{13} M_\odot/h$). These constraints are derived from the \Quijote\ suite of halo catalogues at $z=0$ and $z=1$, each having a $1~(h^{-1}{\rm Gpc})^3$ volume. The solid lines (with triangles) are computed for each primordial shape independently, assuming a fixed cosmology (at fiducial values), while for the dashed-dotted lines we marginalize over the cosmological parameters $\sigma_8$ and $\Omega_m$. This highlights the large degeneracies between the parameters at the level of the halo mass function. For comparison, we also show the corresponding constraints from the power spectrum and bispectrum (horizontal solid and dashed-dotted lines for the independent and joint cases, respectively), as computed previously in \citet{Jung:2022gfa} ($\Mmin = 3.2\times 10^{13} M_\odot/h$). If we consider the unmarginalized HMF results, we see that the $f_{\rm NL}$ constraining power is higher at $z=1$ for the local and equilateral case, despite the smaller number of halos at this redshift; this is clearly due to a stronger response of the HMF to variations in $f_{\rm NL}$ at higher redshift, consistent with previous findings \citep[see, e.g., figure~4 in ][]{Loverde:2008}. The shape is due to the change of sign in the $f_{\rm NL}$ derivative at different masses, discussed in the main text and figure~\ref{fig:deriv-hmf}.} 
    \label{fig:fnl-hmf}
\end{figure}

As a preliminary exercise, in figure~\ref{fig:fnl-hmf}, we show the constraining power of the HMF on the PNG amplitudes $\fNL$ of the three shapes, assuming exactly known cosmological parameters. As expected, the HMF is, in this case, extremely sensitive to the presence of PNG, leading to even tighter constraints than the power spectrum and bispectrum. For example, our Fisher forecast on PNG of the local type is $\sigma(\fNLloc) \sim 30$ at $z=0$, which is in very good agreement with the GNN and deep set results $\sigma(\fNLloc) \simeq 35$ (see section\ \ref{sec:GNN}), and more than twice as small as the equivalent power spectrum $+$ bispectrum forecast error bar.

However, it is well known that there are large degeneracies between $\fNL$ and several cosmological parameters, like $\sigma_8$ or $\Omega_m$ \citep{Maturi:2011am}, as can be verified in figure~\ref{fig:deriv-hmf}.

When we jointly analyze all parameters, these degeneracies increase the errors significantly (by roughly one order of magnitude at $z=1$, and slightly less at $z=0$, where the change of sign of $f_{\rm NL}$ derivative, seen in figure~\ref{fig:deriv-hmf}, helps distinguish it from the response to variations in other cosmological parameters), making them larger than those achievable from the power spectrum and bispectrum combination.

\subsection{Joint constraints with the power spectrum and bispectrum}
\label{sec:joint}

While, as expected, the HMF alone does not produce competitive $f_{\rm NL}$ constraints in comparison with the power spectrum and bispectrum, it does remain interesting to investigate whether a combined analysis of all three statistics can produce significant improvements; this is the main point of the present work. Complementing our previous power spectrum $+$ bispectrum analysis with the HMF can, in principle, benefit us in two ways. First of all, it directly adds extra information about the $f_{\rm NL}$ parameter; also, it could be useful to help break the important degeneracy between $f_{\rm NL}$ and the so-called $b_{\phi}$ bias parameter.

Before presenting our results, let us review and discuss the latter point in more detail. In the presence of local PNG, the halo density fluctuation field $\delta_h(z)$ can be written to leading order as follows \citep{Dalal:2007cu, Matarrese:2008nc, Slosar:2008hx, McDonald:2008sc, Giannantonio:2009ak, Desjacques:2010jw}:
\begin{equation}
    \delta_h(z) = \left[ b_1(z) + \frac{3 \Omega_m H_0^2}{2 D(z) k^2} b_{\phi} f_{\rm NL} \right] \delta_m(z),
    \label{eq:halo_bias}
\end{equation}
where $\delta_m$ is the matter density fluctuation, $D(z)$ is the growth factor and $b_1, b_{\phi}$ are bias parameters, defined respectively as the response of $\delta_h$ to mass density $\delta_m$ and primordial potential $\phi$. It is evident, in this relation, that the scale-dependent signature depends on both $b_{\phi}$ and $f_{\rm NL}$, and that the two parameters are completely degenerate. This issue can be avoided if one assumes, as it was generally done, the universality relation between $b_1$ and $b_{\phi}$, that is,
\begin{equation}\label{eqn:bhiuniv}
b_{\phi} = 2 \delta_c (b_1 - 1) ,
\end{equation}
where $\delta_c$ is the critical density for collapse. However, it was recently pointed out in \citet{Barreira:2020ekm, Barreira:2022sey} that such a relation does not accurately describe the bias of either galaxies, selected by stellar mass, or halos, selected by concentration. Therefore, $b_{\phi}$ is not exactly determined anymore and this reintroduces the $b_{\phi}$-$f_{\rm NL}$ degeneracy problem. To overcome the issue, different studies have focused on using simulations to produce accurate priors on $b_{\phi}$ \citep{Lazeyras:2022koc} and exploiting the multi-tracer technique \citep{Barreira:2023rxn, Sullivan:2023qjr, Karagiannis:2023lsj}. In the present context, the idea is instead to try and break the degeneracy by exploiting the information in the HMF---which selects all halos in each given mass bin---and its direct dependence on $f_{\rm NL}$ and not on $b_{\phi}$.

For clarity, we split the discussion of our results into two parts. Initially, we assume universality in the $b_\phi(b_1)$ relation using eq. (\ref{eqn:bhiuniv}) and we measure the sheer extra information content in the HMF, in the absence of the $b_{\phi}$-$f_{\rm NL}$ degeneracy \footnote{Or, equivalently, we forecast the power spectrum $+$ bispectrum $+$ HMF constraining power on the $b_{\phi} f_{\rm NL}$ parameter combination}. Later on, we instead treat $b_{\phi}$ as a free parameter.

\subsection{Fixing $b_{\phi}$}

\begin{figure}  
    \includegraphics[width=0.99\linewidth]{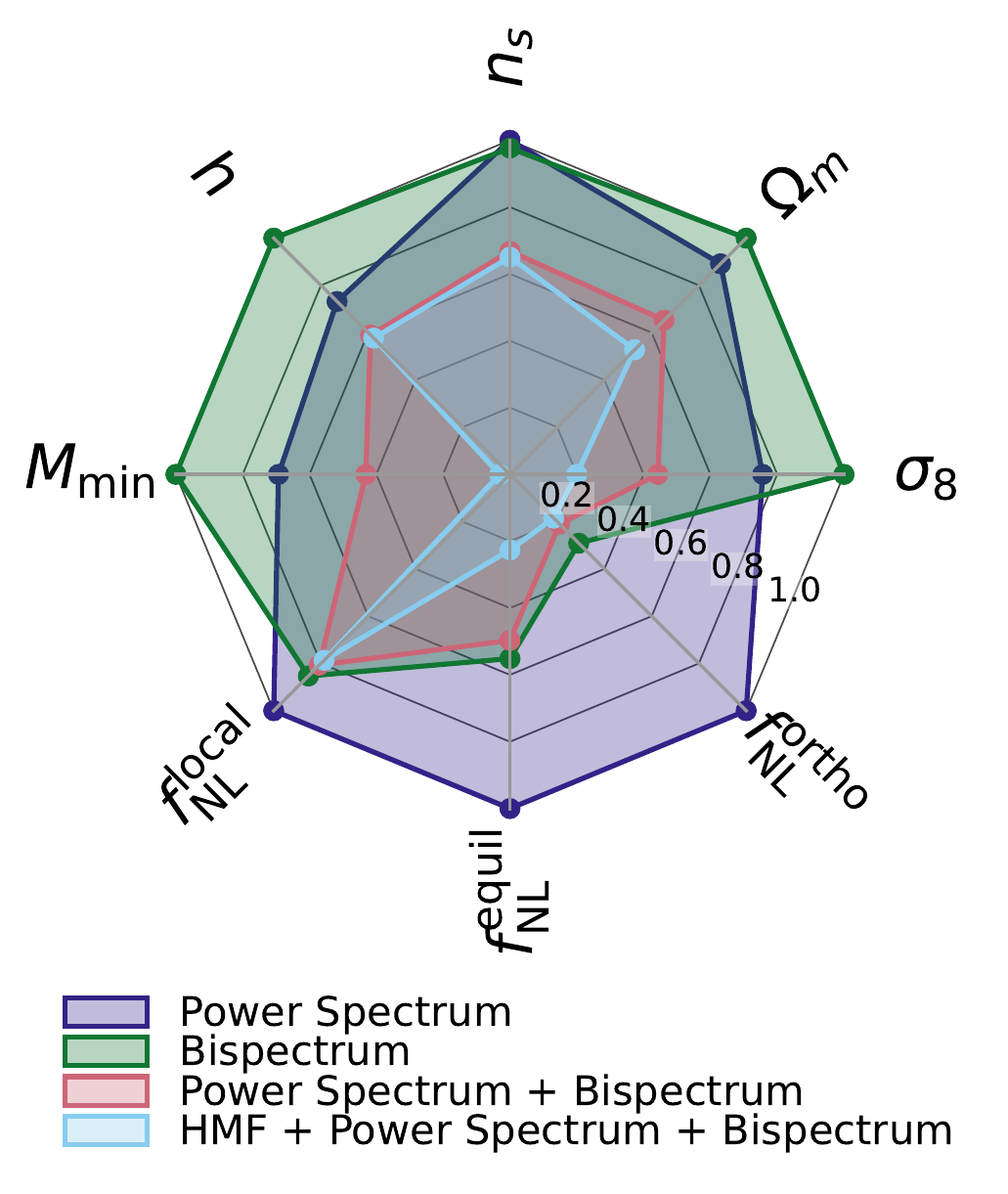}
    \caption{Ratio of $1$-$\sigma$ Fisher error bars on the cosmological parameters and PNG amplitudes from the HMF, power spectrum and bispectrum at $z=1$, assuming $b_\phi$ fixed. This illustrates how including the HMF tightens the constraints on several parameters ($\sigma_8$ and $\fNLeq$ in particular). Note that the values of these error bars are given in table~\ref{tab:results} and figure~\ref{fig:convergence}.}
    \label{fig:errors}
\end{figure}

\begin{figure*}
    \includegraphics[width=0.99\linewidth]{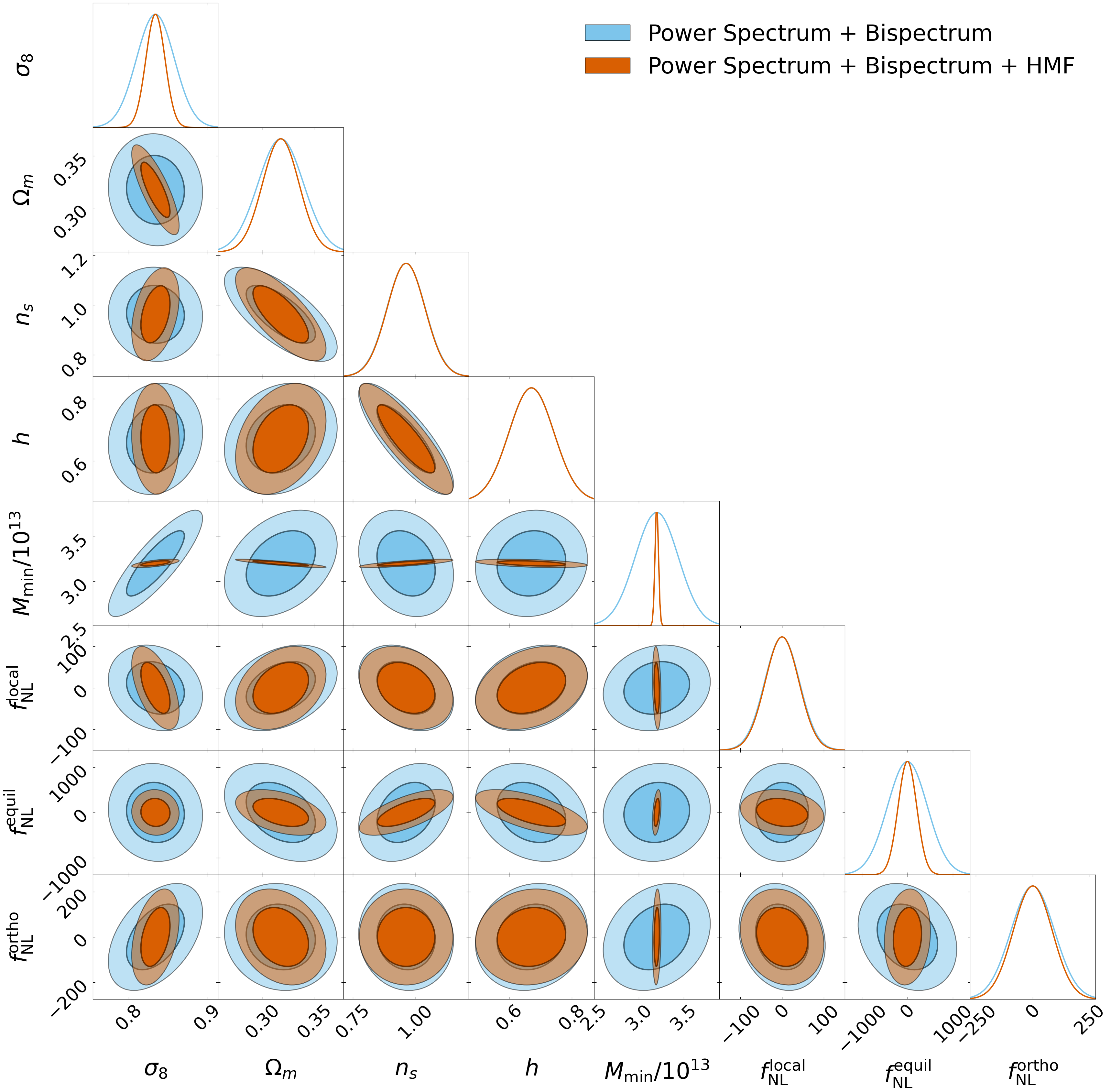}
    \caption{Impact of the HMF on the $1$-$\sigma$ constraints on the cosmological parameters and PNG amplitudes from the halo power spectrum and bispectrum at $z=1$, assuming $b_{\phi}$ is fixed.}
    \label{fig:contours-hmf}
\end{figure*}

The outcome of the first part of the analysis (assuming universality in $b_\phi(b_1)$ is illustrated in figure~\ref{fig:errors} and \ref{fig:contours-hmf} (see also table~\ref{tab:results}). We see that by adding the HMF, the error bars on $\sigma_8$ and $\fNLeq$ become roughly twice as small as the power spectrum $+$ bispectrum result. Moreover, there is also a noticeable improvement for $\Omega_m$ and $\fNLort$. For $\fNLloc$ there is instead no clear improvement; this seems due to the fact that in this case the information content is totally dominated by the power spectrum contribution, via scale dependent bias. Such a contribution is instead smaller for the orthogonal shape and absent for the equilateral case, making the HMF inclusion more important for these scenarios and especially the equilateral one.

Note that we consider only halos with masses above $\sim 4 \times 10^{13} M_\odot/h$ in the HMF, which is larger than the fiducial $\Mmin = 3.2\times 10^{13} M_\odot/h$ used to study the power spectrum and bispectrum. This means that the HMF is not sensitive at all to small variations of $\Mmin$ around the fiducial value. However, through cross-correlated terms with the other summary statistics, the error bars on $\Mmin$ are almost two orders of magnitude smaller\footnote{An important caveat here is that it is important to verify whether this conclusion holds when considering a more complex bias model, which includes higher order bias parameters; this will be done as part of a future work on mock galaxy catalogues, by including numerical derivatives with respect to HOD parameters}. In appendix \ref{app:convergence}, we verify the numerical stability of our results by varying the number of simulations used.

\begin{figure}
    \includegraphics[width=0.99\linewidth]{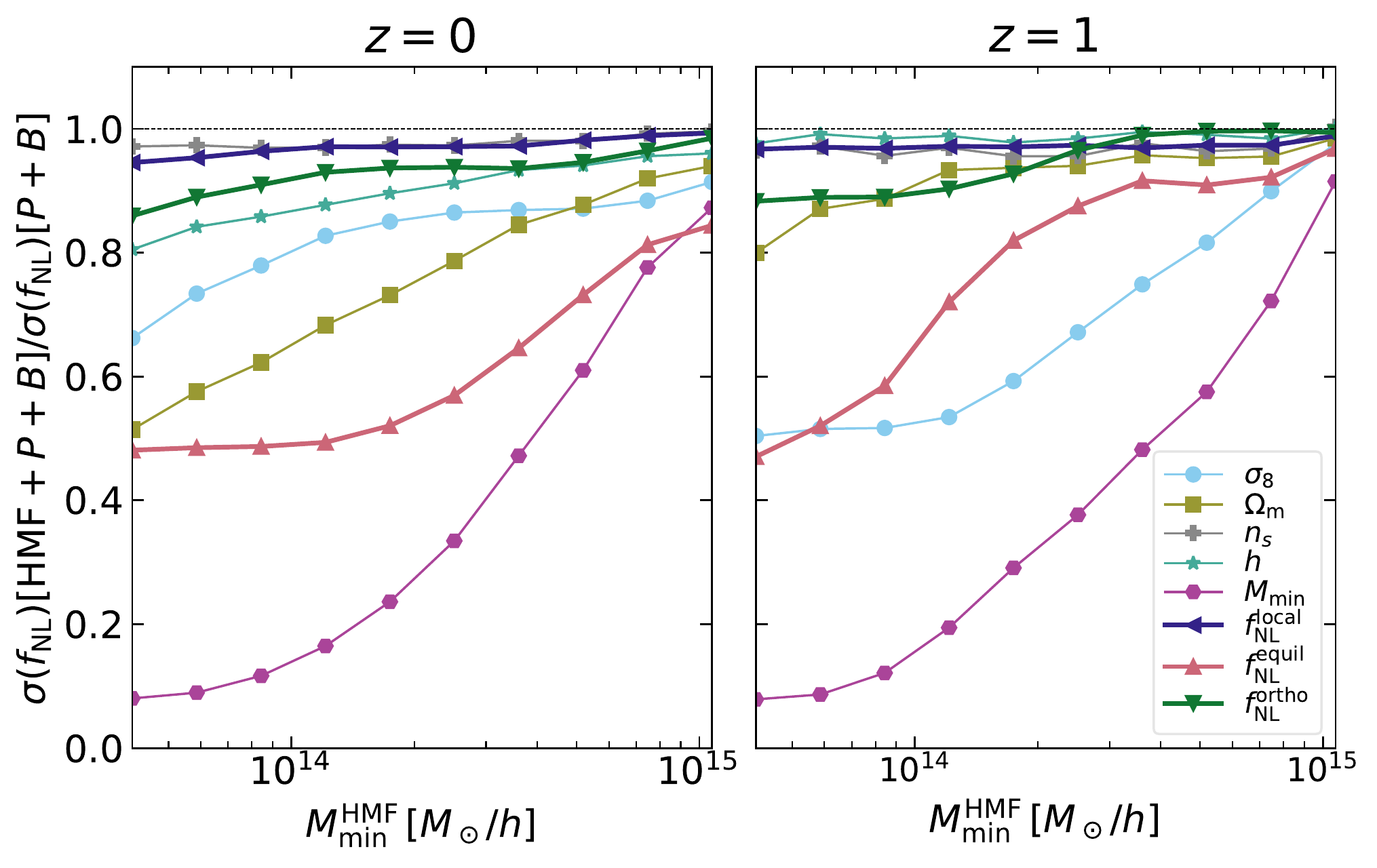}
    \caption{The impact of varying the lowest mass bins of the HMF on the $1$-$\sigma$ Fisher constraints on cosmological parameters and PNG amplitudes from the combination of halo mass function, power spectrum and bispectrum at $z=0$ and $z=1$, assuming $b_{\phi}$ fixed. All errors are normalized by their equivalent using the power spectrum and bispectrum only. Note that we restrict only the mass range for the HMF.}
    \label{fig:hmf_mmin}
\end{figure}

It is interesting to check which halo mass range gives the largest contribution to the observed improvements. To this purpose, we repeat the analysis by varying $\Mmin^\mathrm{HMF}$, the lowest mass bin of the HMF used to evaluate Fisher matrices. Our results are displayed in figure~\ref{fig:hmf_mmin}, which highlights different behaviours for the different parameters considered. Most importantly, for the two PNG parameters $\fNLeq$ and $\fNLort$, halos of intermediate masses ($\sim 2$-$6 \times 10^{14} M_\odot/h$ at $z=0$, and slightly smaller at $z=1$) play a significant role in the observed improvement of constraints, while less massive halos, despite being more numerous, have a much smaller effect. However, the situation is different for cosmological parameters like $\sigma_8$ and $\Om$, where those same less massive halos contain most of the information.

\subsection{Breaking the $b_{\phi}$--$\fNLloc$ degeneracy with the HMF}
\label{sec:bias}

Accounting for the effects of $b_{\phi}$ in our methodology is not straightforward, since $b_{\phi}$ cannot be explicitly included as an input parameter in our simulations and this does not allow us to directly compute the numerical derivative $\partial \mathbf{s} / \partial b_{\phi}$.
To circumvent this issue in a simple way and be able to perform a first test of  the ability of the HMF to remove degeneracies between $b_\phi$ and $\fNLloc$, we then decide here to work under the conservative assumption that these two parameters are fully degenerate at the level of the halo power spectrum and bispectrum. In other words, we assume that
${\partial \mathbf{s} / \partial b_{\phi}} \propto {\partial \mathbf{s} / \partial \fNLloc}$, where $\mathbf{s}$ is either the power spectrum or the bispectrum. For the HMF, we instead set the derivative with respect to $b_\phi$ equal to zero, as it does not depend on this parameter, and compute the $\fNLloc$ derivative as usual.

\begin{figure}
    \includegraphics[width=0.99\linewidth]{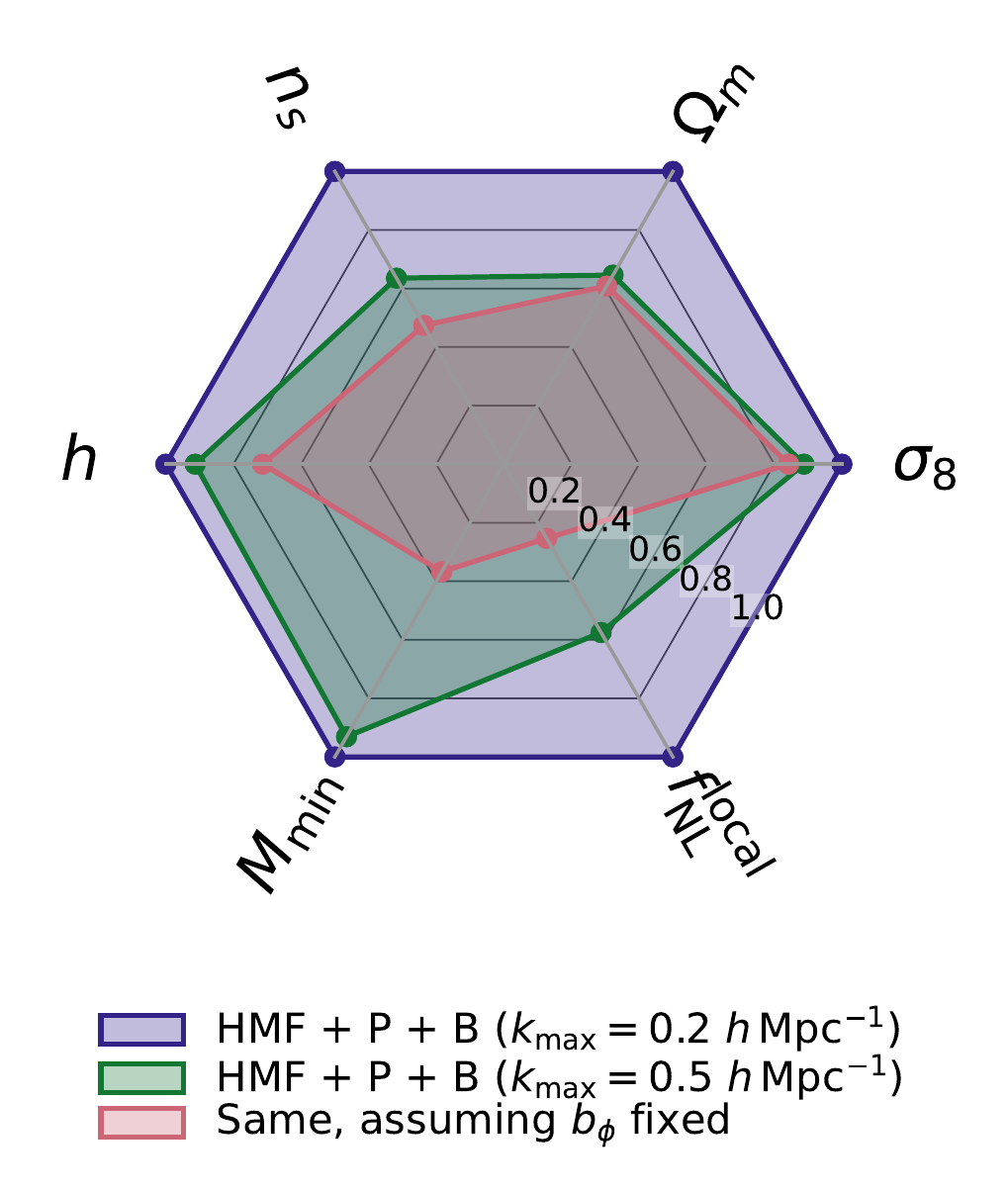}
    \caption{The HMF can break the $b_\phi$-$\fNLloc$ degeneracy in the power spectrum and bispectrum. As in figure\ \ref{fig:errors}, we show normalized $1$-$\sigma$ Fisher error bars derived from the HMF, halo power spectrum and bispectrum at $z=1$. Here we assume that $\fNLloc$ and $b_\phi$ are fully degenerate at the power spectrum and bispectrum level, while the HMF does not depend on $b_\phi$.}
    \label{fig:errors_bphi}
\end{figure}

In figure\ \ref{fig:errors_bphi}, we show the $1$-$\sigma$ Fisher constraints obtained in this assumption and compare them with the ``ideal" ($b_{\phi}$ fixed) constraints derived in the previous section, for different $\kmax$ (see also table~\ref{tab:results}).

The most important result here is that the inclusion of the HMF makes it possible to break the $b_{\phi}$--$\fNLloc$  degeneracy to a level that allows us to produce meaningful $\fNLloc$ constraints without resorting to any prior information on $b_{\phi}$. The final $\fNLloc$ forecast is, however, degraded by a factor of $\sim 2.5$ with respect to the idealized, $b_{\phi}$ fixed case that was shown in figure\ \ref{fig:convergence}. In order to achieve this constraining level it is also crucial to include the information from the power spectrum and bispectrum at non-linear scales ($k$ between $0.2$ and $0.5~\hMpc$), as it helps break degeneracies with several cosmological parameters ($\Om$ in particular).

We corroborate our findings with a simulation-independent analysis based on the halo model \citep[for a review, see][]{Cooray:2002dia, Asgari:2023mej}.
Within this framework, we describe the HMF and halo power spectrum following \citet{Takada:2013jwa}, up to $k_{\rm max} = 0.2 \, h \, {\rm Mpc}^{-1}$. We use the HMF and bias from \citet{2010ApJ...724..878T} using $M_{200,m}$ directly as the mass definition in the mass integration. In the power spectrum analysis of the simulations, the halos are considered point-like, thus we use a Dirac delta as the halo profile. Thanks to the low $\kmax$ we use, the 2-halo term dominates the signal and this approximation is appropriate.
The effect of PNG---here we only consider the local model---is included as a correction to the HMF parametrized according to \citet{LoVerde:2011iz}, and through the scale dependent halo bias shown in equation~\eqref{eq:halo_bias}.
While aware that the $M_{200,m}$ mass does not match the FOF mass used in the rest of the paper, we still consider as observable the HMF divided in 10 bins logarithmically spaced between $3.2 \times 10^{13}$ and  $3.2 \times 10^{15} \, M_\odot/h$.
We bin the halo power spectrum in 30 bins logarithmically spaced between $6.3 \times 10^{-3}$ $\hMpc$ and $0.2$ $\hMpc$. We choose a relatively low $\kmax$ to ensure that non-linearities are negligible at this stage.
In the HMF-halo power spectrum covariance, for which we again follow \citet{Takada:2013jwa}, only the Gaussian terms are included at present.
A more refined analysis, including a wider range of scales and masses, the complete covariance, uncertainties on the parametrization of the HMF and, crucially, the bispectrum will be presented in a future work \citep{Ravennietal}.

\begin{figure}
    \includegraphics[width=0.99\linewidth]{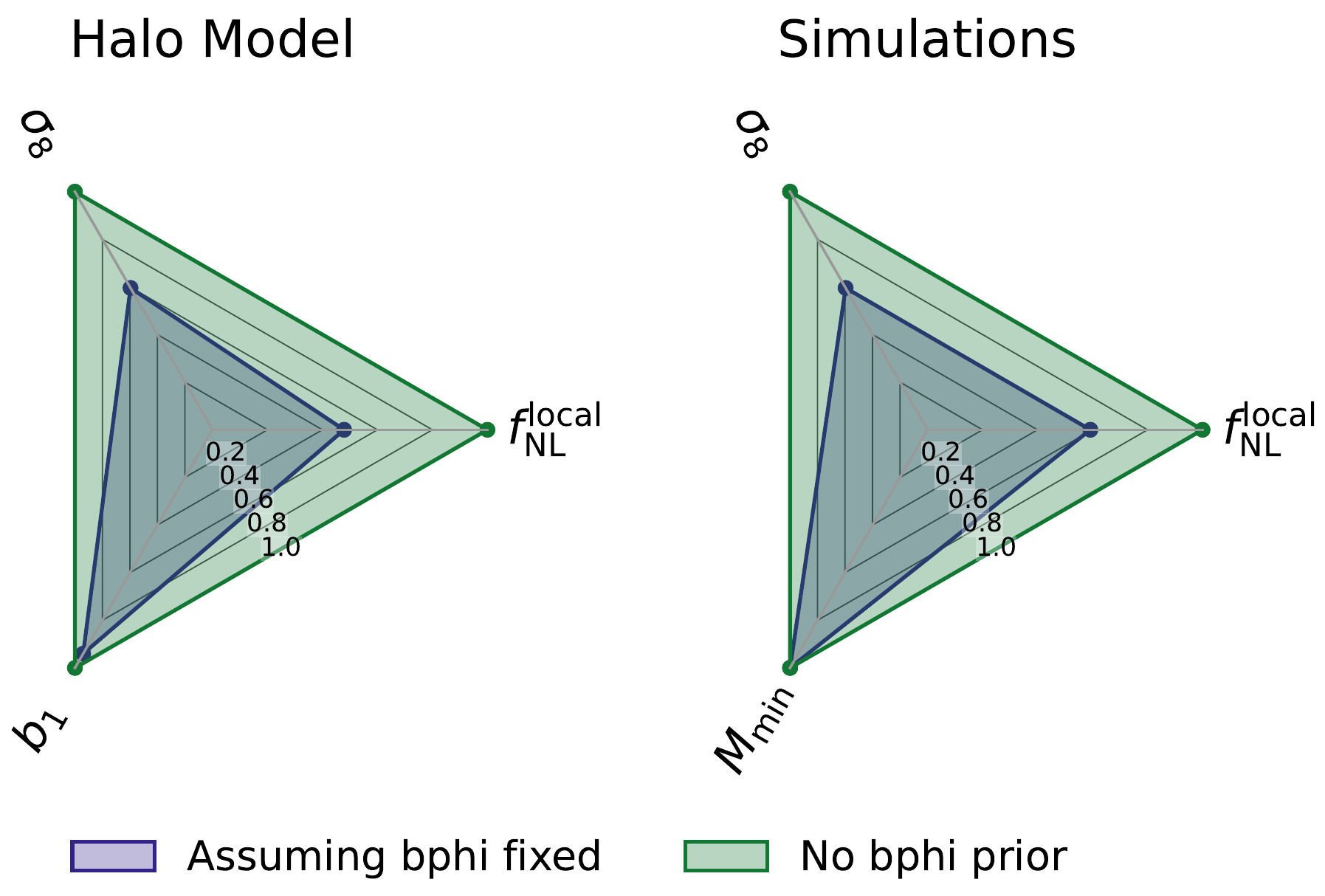}
    \caption{Similar to figure~\ref{fig:errors_bphi}, considering only $\{\sigma_8, \fNLloc\}$ and bias parameters. The $1$-$\sigma$ Fisher constraints include the information contained in the HMF and the power spectrum information up to $\kmax=0.2\,\hMpc$ computed using the halo model on the left, and from simulations on the right. Note that both methods give $\sigma(\fNLloc)\sim 50$ and similar $\sigma(\sigma_8)$ (less than $20\%$ difference).}
    \label{fig:th_vs_sims}
\end{figure}

The results are shown in figure\ \ref{fig:th_vs_sims}, which highlights a very good agreement between our preliminary theoretical computations and the purely simulation-based forecast. This result confirms that a joint analysis including the HMF is an interesting approach that deserves further investigation and could be adopted as a complementary strategy to those already implemented in the literature to address the $b_{\phi}$--$\fNLloc$ degeneracy issue.

\subsection{Removing degeneracies with \Planck\ priors}
\label{sec:planck}

As highlighted in section~\ref{sec:joint}, removing the degeneracies of the HMF using the information from the halo power spectrum and halo bispectrum significantly improves the constraints on PNG of the equilateral type. In this section, we push the idea further by assuming strong but realistic priors on cosmological parameters, based on CMB measurements from \Planck. 

\begin{figure}
    \includegraphics[width=0.99\linewidth]{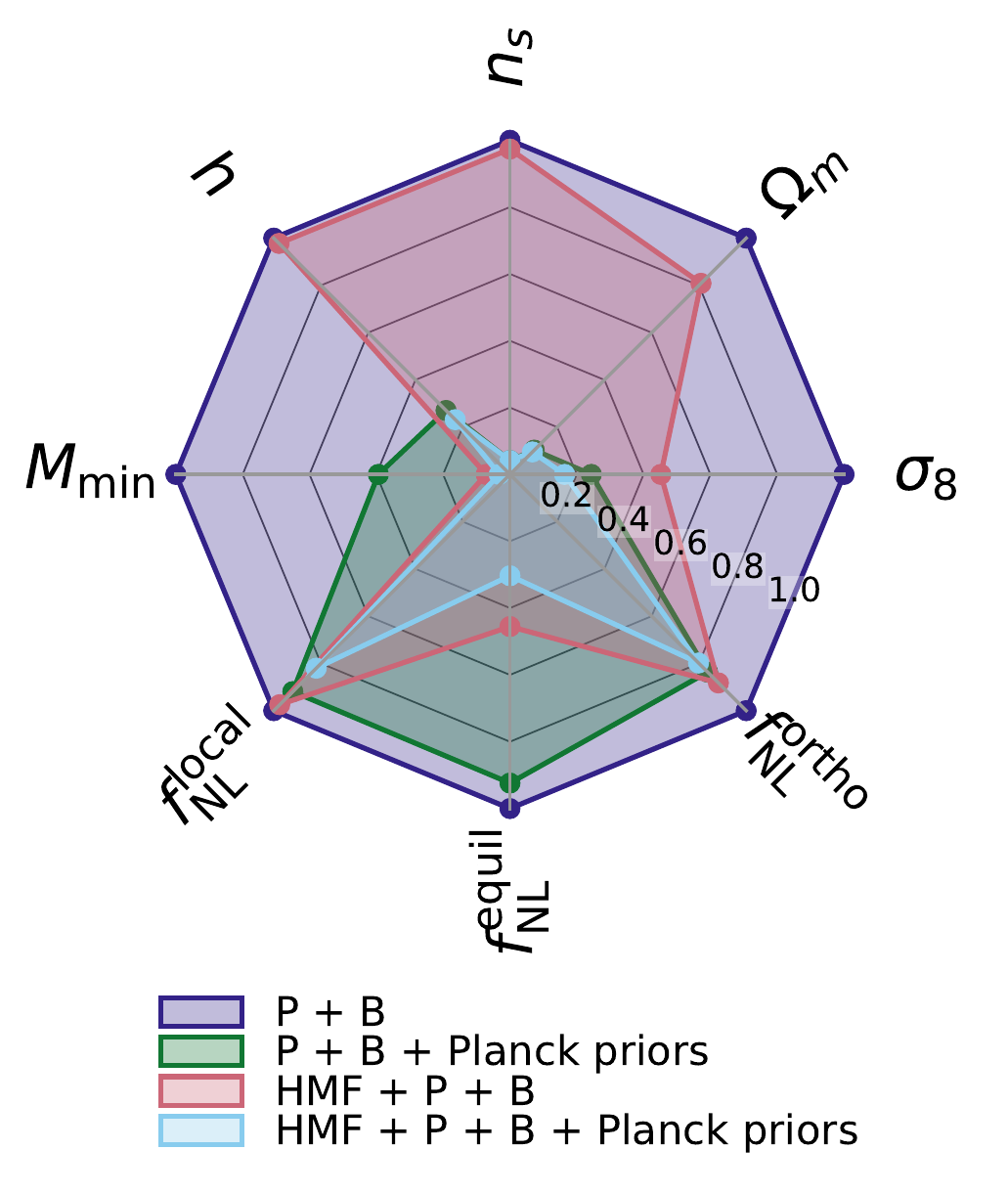}
    \caption{Similar to figure~\ref{fig:errors}, where we include \Planck\ priors on the cosmological parameters $\left\{\sigma_8, \Om, n_s, h \right\}$ and we assume $b_\phi$ fixed.}
    \label{fig:errors_planck}
\end{figure}

We use the same Gaussian likelihood based on the \Planck\ CMB data \citep{Planck:2019nip} as in \citet{Uhlemann:2019gni} in figure~\ref{fig:errors_planck} in addition to our HMF, power spectrum and bispectrum measurements to derive $1$-$\sigma$ Fisher constraints (see also table~\ref{tab:results}). For both $\fNLloc$ and $\fNLeq$ it improves these constraints, while the effect is smaller for $\fNLort$. Note also that the effect is the strongest when the HMF is also considered in the analysis, meaning it removes degeneracies between the PNG and cosmological parameters at the level of the HMF. Concerning numerical convergence with the number of simulations used to compute the derivatives, including these \Planck\ priors also improves it significantly, where only $\fNLeq$ is not optimally constrained for the power spectrum + bispectrum case, and all parameters have converged when we add the HMF information.

\begin{deluxetable}{c|c|c|c}[]
\tablecaption{The $1$-$\sigma$ constraints on cosmological parameters and PNG amplitudes at $z=1$ obtained by combining the information of the halo power spectrum, bispectrum and mass function, each measured from the \Quijote~and \QuijotePNG~simulations.  \label{tab:results}}
\tablehead{& $b_\phi$ fixed & No prior on $b_\phi$ & \Planck\ priors}
\startdata 
$\sigma_8$ & $0.012$  & $0.013$ & $0.005$ \\
$\Omega_m$ & $0.018$ & $0.017$ & $0.002$ \\
$n_s$ & $0.075$ & $0.075$ & $0.003$ \\
$h$ & $0.072$ & $0.071$ & $0.017$ \\
\hline
$\fNLloc$ & $40$ & $89$ & $34$ \\
$\fNLeq$ & $203$ & & $136$ \\
$\fNLort$ & $85$ & & $79$ \\
\hline
$\Mmin/10^{13}$ & $0.019$ & $0.045$ & $0.009$\\
\enddata
\end{deluxetable}

\section{Conclusion}
\label{sec:conclusion}

In this work, we presented a combined analysis of the power spectrum, bispectrum, and mass function of dark matter halos in the \QuijotePNG\ simulation suite. Our main goal was to verify whether adding the HMF to our previous joint power spectrum and bispectrum analyses \citep{Coulton:2022qbc,Jung:2022rtn,Coulton:2022rir,Jung:2022gfa} could lead to improved constraints on primordial non-Gaussianity.
The main underlying reason behind this analysis is that the HMF turned out to be the statistics used by a sophisticated graph neural network when carrying out 
a preliminary field-level likelihood-free inference calculation. Furthermore, the HMF tail has been known for a long time to be strongly sensitive to PNG. Finally, the HMF not only carries complementary information to the power spectrum and bispectrum, but also does not suffer from the $b_{\phi}$--$\fNLloc$, assembly bias-PNG degeneracy that was recently pointed out in \citet{Barreira:2020ekm, Barreira:2022sey} as an important issue in the analysis of local PNG.

Our results show that the HMF can indeed play a significant role in tightening the expected PNG bounds and breaking parameter degeneracies when its contribution is added to those of the power spectrum and bispectrum.
In the first part of our analysis, we remove a priori the $b_{\phi}$--$\fNLloc$ degeneracy by assuming universality in the $b_{\phi}(b_1)$ relation; i.e, we set $b_{\phi} = 2 \delta_c (b_1 - 1)$. In this case, we see that the HMF is able to improve equilateral $f_{\rm NL}$ constraints by roughly a factor $2$ and orthogonal $f_{\rm NL}$ constraints by $10\%$. Constraints on PNG of the local type are instead unchanged, since in this idealized scenario the local PNG information is dominated by the large scale power spectrum modes, via scale dependent bias.

In the second part of the analysis, we instead treat $b_{\phi}$ as a free parameter and assume that the responses of the halo power spectrum and bispectrum to changes in $b_{\phi}$ and $\fNLloc$ are identical; that is, we assume that these two parameters are fully degenerate in a joint analysis of the power spectrum and bispectrum. Starting with this setup, we then see that the additional inclusion of the HMF is able to break the $b_{\phi}$-${\fNLloc}$ degeneracy at a significant level, without the need to rely on any prior on $b_{\phi}$ or any other external information. More precisely, our final $\fNLloc$ constraints after marginalizing over $b_{\phi}$ and other standard cosmological parameters are now degraded by a factor $\sim 2.5$, compared to the ideal case in which $b_{\phi}$ is fixed by the universality relation. We confirmed these results with a semi-analytical, halo model based evaluation of the Fisher matrix, in which we restrict ourselves to the power spectrum and HMF, after verifying that for local PNG these two observables give the dominant contributions to the final sensitivity. We note that to achieve the claimed level of precision on $\fNLloc$, it is important to include non-linear scales in the analysis, up to $k_{\rm max} = 0.5$ $\hMpc$ since they help break additional important degeneracies that affect the HMF constraining power. We also stress that \QuijotePNG\ simulations have a cosmological volume of $1$  $(h^{-1}{\rm Gpc})^{3}$, making it not straightforward to generalize our forecasts to, e.g., a Euclid-like or other coming survey settings. For the same reason, a direct comparison with other forecasts---such as those based on the multi-tracer methodology and placing suitable priors on $b_{\phi}$---is not easy to make at the moment. In a forthcoming publication, \citet{Ravennietal}, we will produce more detailed semi-analytical predictions for future surveys based on the halo model. 

The results presented here have to be considered as preliminary also, as they rely on a simplified bias model for our tracers, and they do not account for systematic effects in the determination of the HMF from actual observations. Indeed, the dark matter mass of a halo is a quantity that is notoriously difficult to measure observationally, especially for high-redshift objects. Halos are complex and dynamic structures that are almost exclusively probed by the signal broadcast by the baryons they host. (Dark) Mass measurements tend to require sophisticated and labor-intensive observations, which is unfeasible for a large number of objects, as needed for the HMF. Moreover, the sample completeness (for the host halo, not the tracers) needs to be known exquisitely well, which may constitute a formidable challenge.
Among the most promising approaches are the Sunyaev Zeldovich effect-selected clusters (signal at mm wavelengths) \citep{Mroczkowski:2018nrv}, X-ray clusters \citep{Pratt:2019cnf} and (optical) gravitational lensing mass determination \cite[e.g.][]{Murray:2022iyi}. For example, cluster catalogs will increase drastically with a suite of forthcoming experiments: eROSITA \citep{eROSITA:2020emt}, Simons Observatory \citep{SimonsObservatory:2018koc}, Euclid \citep{2011arXiv1110.3193L}, Roman \citep{Akeson:2019biv} and Rubin \citep{LSST:2008ijt}. Cluster masses will not be measured directly but inferred through proxies; these proxies, however, will be provided as a product of these surveys, and are expected to be or be made robust and reliable. An important ingredient for any HMF analysis would be to robustly quantify the probability distribution of the proxies as a function of the true halo mass. This can then be simply folded into the error budget and the uncertainty propagated through to the inferred parameters. 

The results shown in this paper clearly show that a joint analysis of the HMF, power spectrum and bispectrum of LSS tracers is a promising approach to constrain PNG, hence providing another motivation for further investigation in this direction and for addressing the aforementioned observational issues.

\section*{Acknowledgements}
\noindent
GJ acknowledges support from the ANR LOCALIZATION project,
grant ANR-21-CE31-0019 / 490702358 of the French Agence Nationale de la Recherche. AR acknowledges support from PRIN-MIUR~2020 METE, under contract no. 2020KB33TP. The work of FVN is supported by the Simons Foundation. DK is supported by the South African Radio Astronomy Observatory and the National Research Foundation (Grant No. 75415). LV acknowledges “Center of Excellence Maria de Maeztu 2020-2023” award to the ICCUB (CEX2019-000918-M funded by MCIN/AEI/10.13039/501100011033).

\bibliographystyle{aasjournal}
\bibliography{biblio}

\appendix

\section{Examination of the initial conditions}
\label{app:IC-tests} 

The procedure used to generate the simulation initial conditions (ICs) in \citet{Coulton:2022qbc} is designed to produce a specific bispectrum. However, the method additionally modifies all other N-point functions. The most well studied byproduct of this procedure is modifications to the power spectrum. \citet{Scoccimarro:2011pz} showed that it must be taken in when choosing how to generate the ICs to avoid having corrections that dominate the power spectrum. In \citet{Coulton:2022qbc,Jung:2022rtn}, the ICs were validated by examining the power spectrum and bispectrum. Those tests showed that the modifications to the power spectrum are small and the correct bispectrum was generated. A concern for the results presented in this work, and other studies of statistics beyond the 2- and 3-point functions, is that the ICs may have unphysically large higher order N-point functions that impact the results. The power spectrum and the trispectrum are the leading order unwanted byproducts of the IC generation procedure. If we can show that corrections to both are small, it is reasonable to assume that the impact of the unphysical higher N-point functions of the ICs are negligible for studies of the halo mass function and other statistics of the simulations. Given that the power spectrum has already been validated, in this Appendix we present an investigation into the properties of the trispectrum.

\subsection{Trispectrum estimation}

The trispectrum is defined as
\begin{align}
\langle \delta(\mathbf{k}_1) \delta(\mathbf{k}_2) \delta(\mathbf{k}_3) \delta(\mathbf{k}_4) \rangle  = T(k_1,k_2,k_3,k_4,K_a,K_b),
\end{align}
where $k_i=|\mathbf{k}_i|$, $K_a = | \mathbf{k}_1+\mathbf{k}_2|$ and $K_b = | \mathbf{k}_1+\mathbf{k}_3|$. Estimating the full trispectrum is computationally highly challenging so, in this work, we measure trispectra averaged over $K_b$, i.e.
\begin{equation}
    \mathcal{T}(k_1,k_2,k_3,k_4,K_a) \propto \sum\limits_{K_b} T(k_1,k_2,k_3,k_4,K_a,K_b).
\end{equation}
A binned version of this can be estimated as
\begin{align}
  &  \hat{\mathcal{T}}(k_a,k_b,k_c,k_d,K_E) =\frac{1}{N_{a,b,c,d,E}} \int\prod_{i=1,4}\frac{\mathrm{d}^3k_i}{(2\pi)^3} \int\frac{\mathrm{d}^3 K_a}{(2\pi)^3}\nonumber  \\  & (2\pi)^3\delta^{(3)}(\mathbf{k}_1+\mathbf{k}_2-\mathbf{K}_a) (2\pi)^3\delta^{(3)}(\mathbf{K}_a-\mathbf{k}_3-\mathbf{k}_4)  W_a(\mathbf{k}_1)\nonumber \\ & W_b(\mathbf{k}_2)W_c(\mathbf{k}_3) W_d(\mathbf{k}_4) W_E(\mathbf{K}_a)\delta(\mathbf{k}_1) \delta(\mathbf{k}_2) \delta(\mathbf{k}_3) \delta(\mathbf{k}_4),
\end{align}
where $W_a(k)$ selects modes that lie within binned $a$ and $N_{a,b,c,d,E}$ is the normalization. In this work, we use 14 equally spaced bins between $k=0.0102$ $h$/Mpc to $k=0.193$ $h$/Mpc. By utilizing 
\begin{equation}
    \delta^{(3)}(\mathbf{k_1}+\mathbf{k_2}+\mathbf{k_3}) = \int\mathrm{d}^3{x}e^{i\mathbf{x}\cdot\mathbf{k})} \delta(\mathbf{k}),
\end{equation}
we efficiently implement the estimator by first computing
\begin{equation}
    \delta_{W_a}(\mathbf{x}) = \int\frac{\mathrm{d}^3 k_a}{(2\pi)^3} e^{i\mathbf{x}\cdot{k}}\delta(\mathbf{k})W_a(\mathbf{k}),
\end{equation}
then computing
\begin{align}
    D_{ab}(\mathbf{K})= \int\mathrm{d}^3x e^{-i\mathbf{x}\cdot\mathbf{K}} \delta_{W_a}(\mathbf{x})\delta_{W_b}(\mathbf{x})
\end{align}
and then the estimate is given by
\begin{align}
    &  \hat{\mathcal{T}}(k_a,k_b,k_c,k_d,K_E) =\nonumber \\ & \frac{1}{N_{a,b,c,d,E}} \int\frac{\mathrm{d}^3 K}{(2\pi)^3}D_{ab}(\mathbf{K}) D_{cd}(-\mathbf{K})W_E(\mathbf{K}).
\end{align}
The normalization is obtained by evaluating this estimator (without the $N_{a,b,c,d,E}$ term) on maps with $\delta(\mathbf{k})=1$.

\subsection{Trispectrum of the initial conditions}

To perform a stringent test of the ICs, we study the difference between the trispectrum of the initial conditions with $f_\mathrm{NL}\neq 0$ and $f_\mathrm{NL}= 0$ i.e
\begin{align}
  &  \hat{\mathcal{T}}^\mathrm{diff}(k_a,k_b,k_c,k_d,K_E)= \nonumber \\&\hat{\mathcal{T}}^{f_\mathrm{NL}\neq0}(k_a,k_b,k_c,k_d,K_E)-\hat{\mathcal{T}}^{f_\mathrm{NL}=0}(k_a,k_b,k_c,k_d,K_E).
\end{align}
This cancels the leading noise contribution to the trispectrum measurement. 

\begin{figure}
    \includegraphics[width=0.99\linewidth]{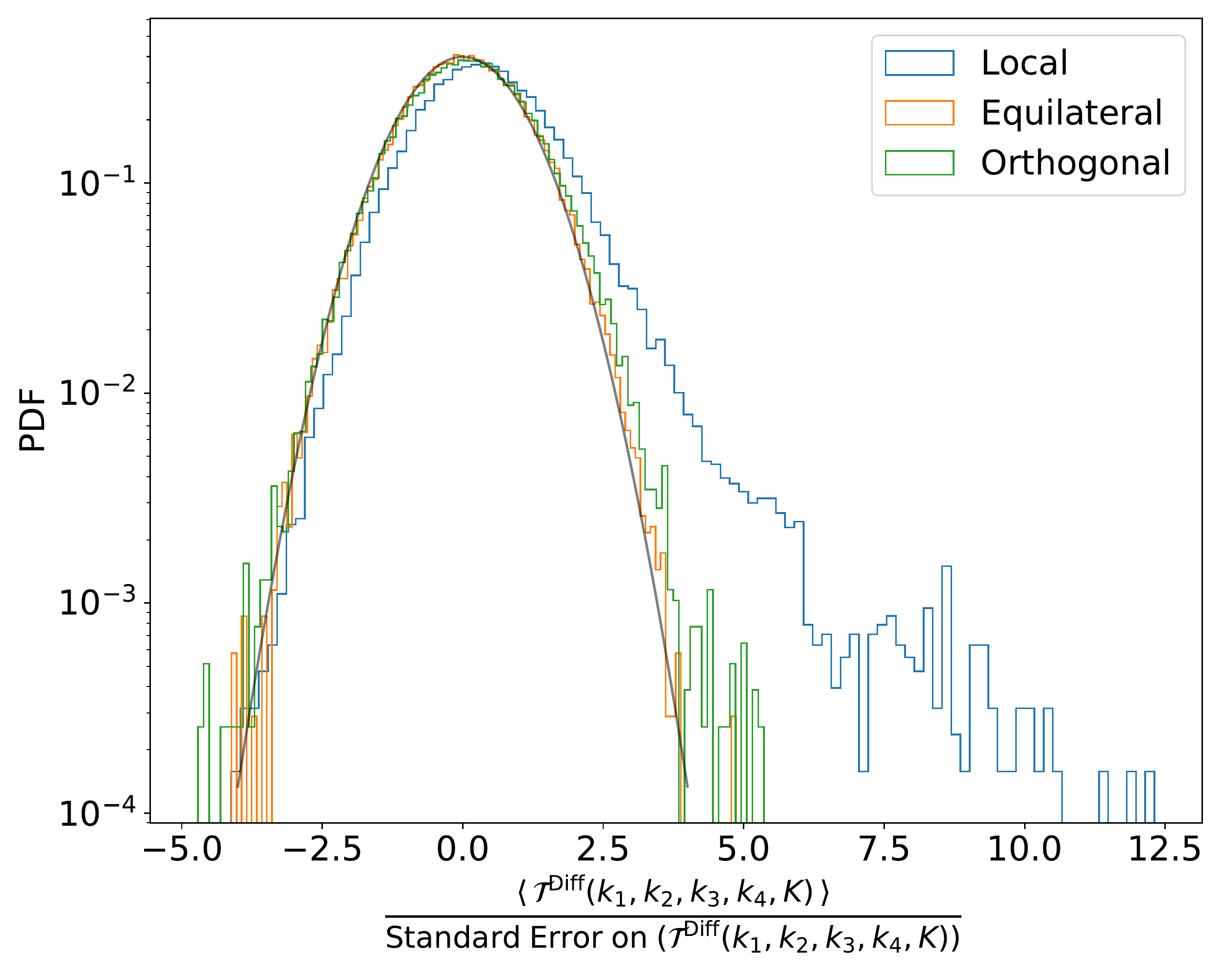}
    \caption{Significance of the detection of the trispectrum in the initial conditions for the three types of primordial non-Gaussianity. This is computed using 200 simulations of each type of primordial non-Gaussianity}
    \label{fig:trispectrumICs}
\end{figure}

The results are shown in figure~\ref{fig:trispectrumICs}. For equilateral, there is no detectable trispectrum. For orthogonal non-Gaussianity, there are small hints of a trispectrum signal. As this measurement uses 200 simulations and a method to cancel the cosmic variance, it is likely that this small trispectrum is negligible. However, the local case shows significant evidence of a trispectrum. This is not unexpected. Local primordial non-Gaussianity is generated in these simulations by
\begin{align}
    \Phi(\mathbf{x})= \phi^G(\mathbf{x})+f_\mathrm{NL}\left(\phi^G(\mathbf{x})^2-\langle(\phi^G(\mathbf{x})\rangle \right)
\end{align}
where $\Phi^G(\mathbf{x})$ is the Gaussian primordial potential potential. This generates a primordial trispectrum known as $\tau_\mathrm{NL}$ \citep{Kogo_20006}. In many inflationary models, $\tau_\mathrm{NL}$ is generated with local non-Gaussianity and thus, the trispectrum seen here is physical.

These trispectra measurements suggest that unphysical higher order N-point functions are not significant in our simulations.

\section{Analyses at other redshifts}
\label{app:other-redshifts}

\begin{figure*}
    \includegraphics[width=0.99\linewidth]{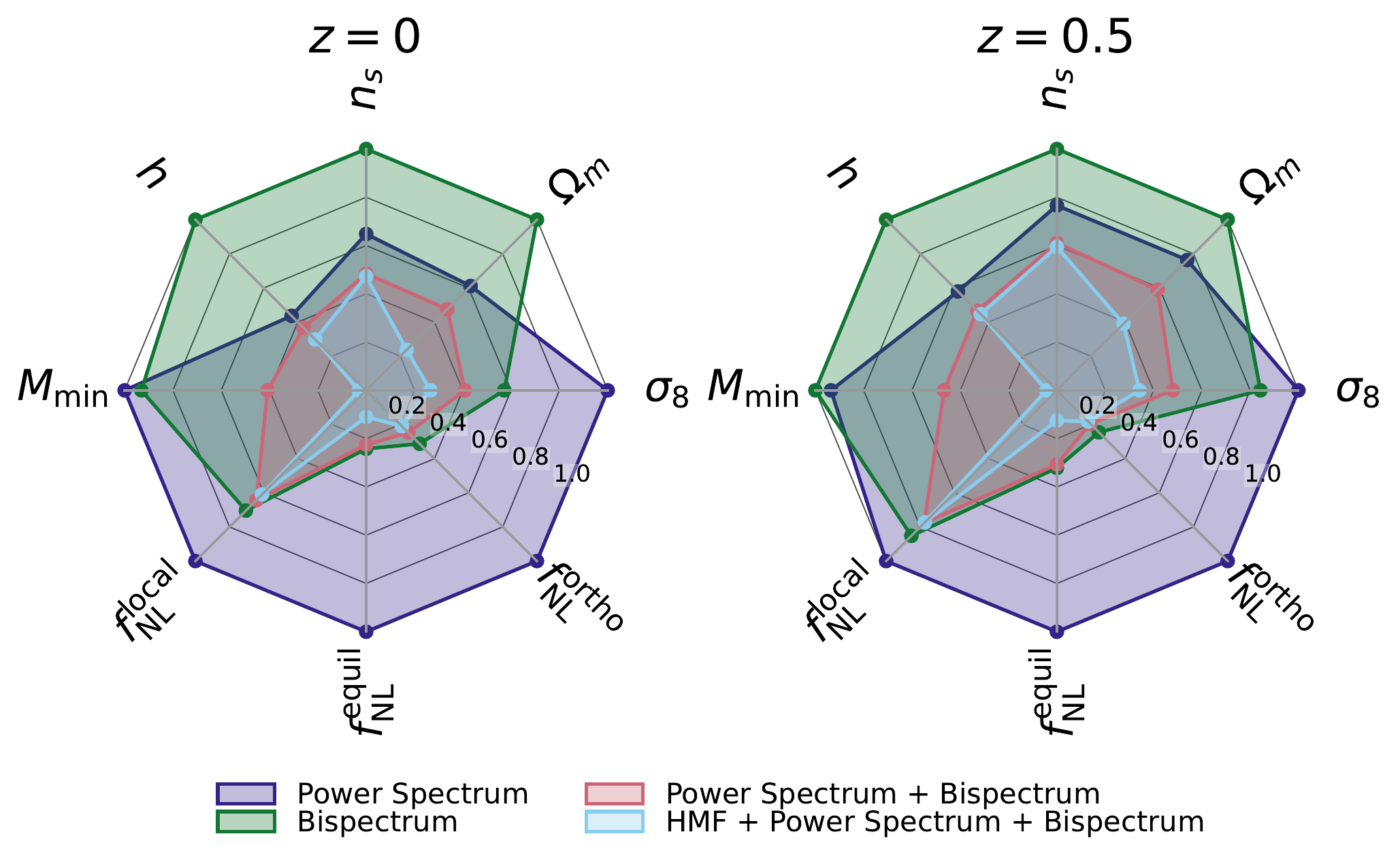}
    \caption{Similar to figure \ref{fig:errors}, at redshifts $z=0$ and $0.5$ and with $b_\phi$ fixed.}
    \label{fig:errors_other_redshifts}
\end{figure*}

We have performed a similar analysis using the \Quijote\ snapshots at $z=0.5$ and $0$ to verify that our conclusions hold at other lower redshifts. As can be seen in figure\ \ref{fig:errors_other_redshifts}, this is indeed the case. For all parameters, the relative improvements due to including the HMF in the Fisher analysis are of the same order (note, however, that the difference between the halo power spectrum and bispectrum results is more pronounced at lower redshifts).

\section{Convergence of numerical derivatives}
\label{app:convergence}

\begin{figure*}
    \includegraphics[width=0.49\linewidth]{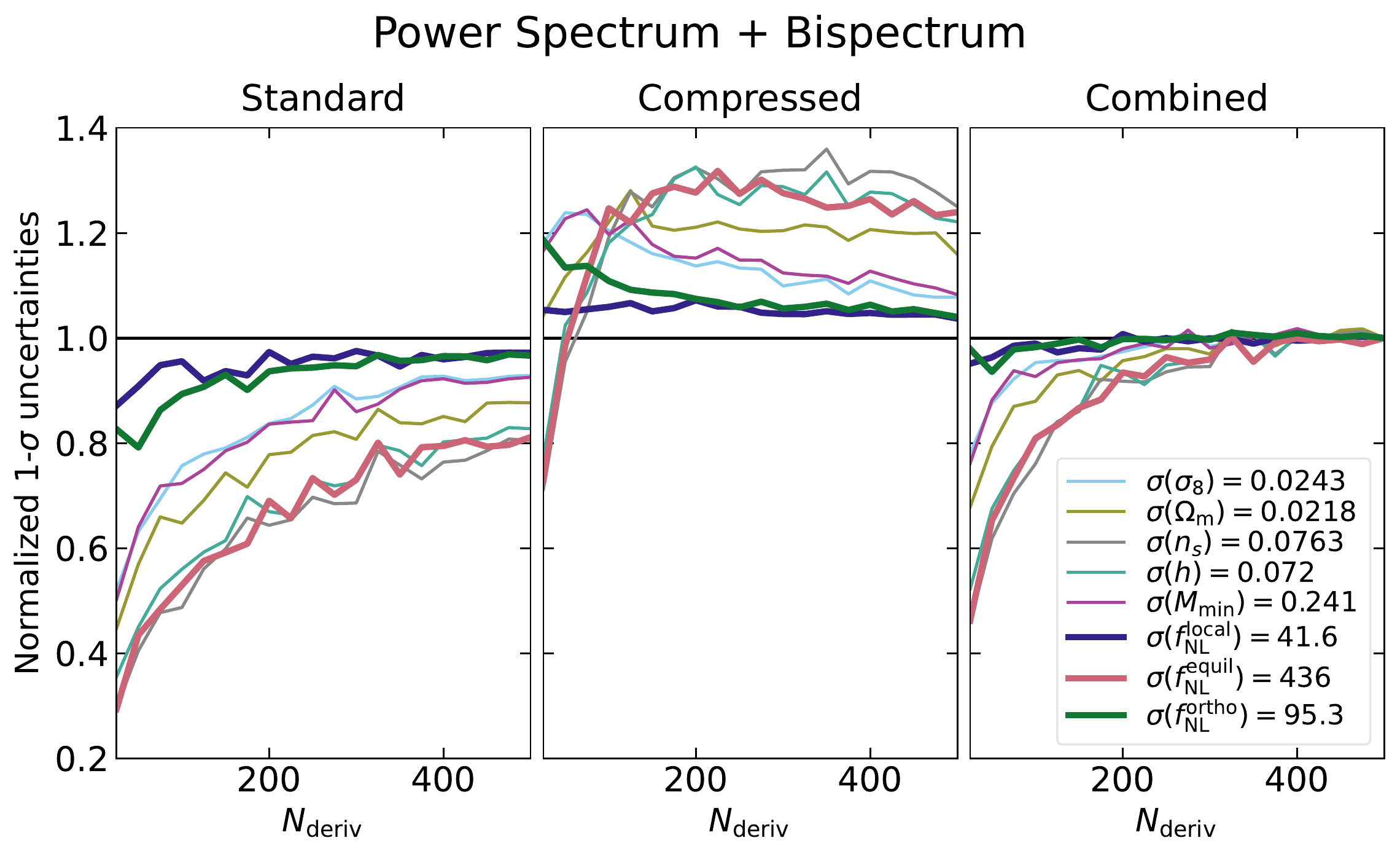}
    \includegraphics[width=0.49\linewidth]{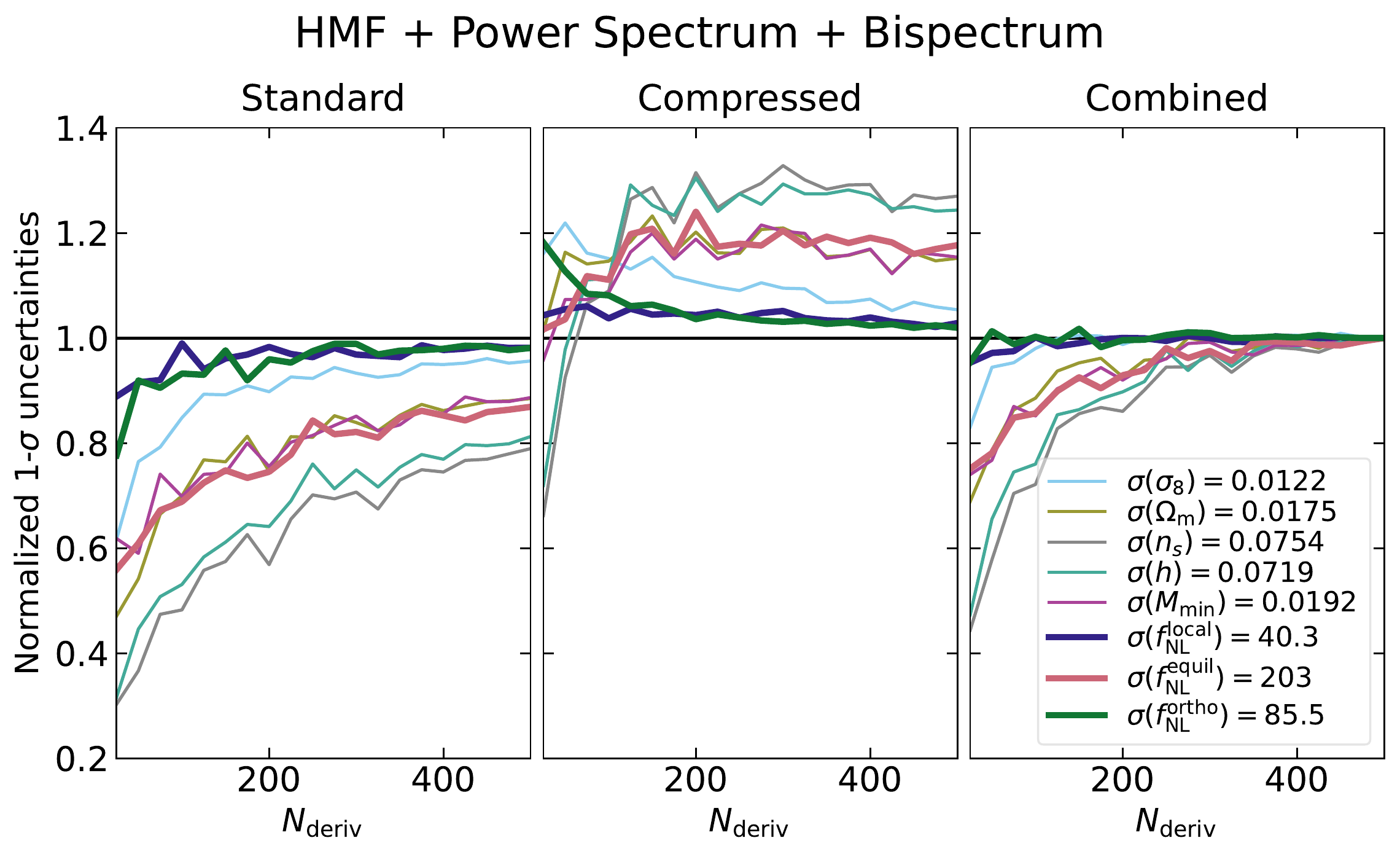}
    \caption{Stability of the Fisher $1$-$\sigma$ error bars when varying the number of simulations used to compute derivatives for the three methods described in section~\ref{sec:fisher} (standard, compressed and combined). In the left panels, the analysis includes the power spectrum (monopole + quadrupole) and bispectrum (monopole) information of the halo field at $z=1$, with scales up to $\kmax=0.5~\hMpc$. In the right panels, we also consider the HMF (for halos with a mass larger than $4.1\times 10^{13} M_\odot/h$). All error bars are normalized by their respective combined Fisher results, given explicitly in the legend for all parameters. They show that adding HMF can significantly reduce the error bars, in addition to improving the numerical convergence of the results (smaller relative differences between the compressed and standard methods) for several parameters, in particular $\sigma_8$ and $\fNLeq$. Note that the lines corresponding to PNG parameters are in bold.}
    \label{fig:convergence}
\end{figure*}

In figure~\ref{fig:convergence}, we study the impact of varying the number of simulations used to compute numerical derivatives on the $1$-$\sigma$ Fisher constraints, both with and without including the HMF in the analyses. This shows that the parameters for which the improvement due to the HMF is the largest (i.e.\ $\sigma_8$ and $\fNLeq$) also have a better numerical convergence with the number of simulations (a smaller difference between the standard and conservative compressed Fisher methods). Note also the stability of the combined Fisher results (variations at the $\%$ level) when using more than $200$ simulations for the derivatives.

\end{document}